\documentclass{aa}

\usepackage{graphicx}
\usepackage{txfonts}
\usepackage{xcolor}
\usepackage{natbib}
\usepackage{hyperref}
\begin{document} 

\defcitealias{2026A&A...709A..21S}{VMS26}

   \title{A probabilistic approach to separate mergers and jellyfish galaxies:}

   \subtitle{Application of morphological non-parametric indexes to the DESI Legacy Survey}

   \author{V. M. Sampaio
          \inst{1,2,}
          \and
          Y. Jaffé\inst{1,2}
          \and
          A. Moretti\inst{3}
          \and
          C. Lima-Dias\inst{4}
          \and
          H. Méndez-Hernández\inst{4}
          \and
          S. V\'eliz Astudillo\inst{4}
          \and
          A. Monachesi\inst{4}
          \and
          S. Torres-Flores\inst{4}
          \and
          A. Krabbe\inst{5}
          \and 
          C. Dulcien\inst{6}}

   \institute{Departamento de Física, Universidad Técnica Federico Santa María, Av. España 1680, Valparaíso, Chile\\
              \email{vitorms999@gmail.com}
         \and
             Millennium Nucleus for Galaxies (MINGAL), Chile\\
         \and
             INAF—Astronomical Observatory of Padova, Vicolo dell'Osservatorio 5, I-35122 Padova, Italy\\
         \and
             Departamento de Astronomía, Universidad de La Serena, Avda. Raúl Bitrán 1305, La Serena, Chile\\
         \and
             Universidade de São Paulo, IAG, Rua do Matão 1226, Cidade Universitária, São Paulo 05508-900,Brazil\\
         \and
             Departamento de Astronomía, Facultad Ciencias Físicas y Matemáticas, Universidad de Concepción, Av. Esteban Iturra s/n Barrio Universitario, Casilla 160, Concepción, Chile\\
             }

\date{Received XXXX; accepted YYYY, in original form ZZZZ}

\abstract
    {Galaxy morphology is a key tracer of galaxy evolution because it reflects the structural transformations produced by internal and external physical processes. However, non-parametric morphological measurements depend on image depth, spatial resolution, segmentation, and preprocessing, and diagnostic boundaries calibrated on one dataset may therefore not be directly applicable to other imaging surveys.}
    {We investigate how non-parametric morphology can be used to classify galaxies in the DESI Legacy Imaging Surveys and test whether ram-pressure-stripping candidates can be distinguished from merger-driven disturbances using imaging-based structural information.}
    {We use \texttt{galmex} measurements of ten non-parametric indices for reference samples of regular spirals, regular ellipticals, mergers, and jellyfish galaxies selected from Galaxy Zoo 1, Galaxy Zoo DECaLS, and \textit{Fishing for Jellyfish Galaxies}. We rank all two-dimensional combinations of indices using a cross-validated Bayesian classifier based on kernel density estimation and combine the full set of indices using calibrated LightGBM models.}
    {The best-performing planes for separating regular ellipticals, regular spirals, and perturbed galaxies combine Shannon entropy with $M_{20}$ or asymmetry-related indices, reaching balanced accuracies of $0.76$--$0.77$. We use the $E$-$M_{20}$ and $E$-$A_3$ planes to define diagnostic regions calibrated specifically for Legacy imaging. Distinguishing mergers from jellyfish galaxies is more difficult: concentration appears in all top-ranked combinations, and the best two-dimensional planes reach balanced accuracies of approximately $0.73$. The calibrated three-class LightGBM model reaches a balanced accuracy and macro-averaged F1-score of $0.89$, while the four-class model reaches values of $0.75$ and $0.76$, respectively. Regular galaxies and jellyfish candidates are recovered reliably, whereas mergers remain the most ambiguous class. We provide the resulting calibrated class probabilities and preferred classifications for $1,744,454$ galaxies in the DESI Legacy Imaging Surveys.}
    {Non-parametric morphology efficiently identifies broad morphological disturbance in wide-field imaging, but determining the physical origin of that disturbance is more challenging. Jellyfish candidates occupy a more coherent region of structural parameter space than mergers, whose diverse appearances lead to substantial overlap with the other morphological classes.}
   \titlerunning{Non-parametric morphology in Legacy}
   \authorrunning{Sampaio et al.(2026b)}

   \maketitle

\section{Introduction}

Galaxy morphology is a key observable of galaxy formation and evolution because it traces the spatial distribution of stars and gas and reflects the dynamical state of a galaxy \citep[e.g.,][]{1969ApJ...155..393P,2015ApJ...812...29T}. Morphology changes as galaxies accrete gas, interact with companions, and respond to their environment. These transformations underpin well-established relations between morphology, environment, colour, stellar mass, and star-formation history \citep[e.g.,][]{Dressler,Strateva1,2012MNRAS.424..232W,2024MNRAS.532..982S}. Quantitative morphology therefore provides a scalable way to investigate the physical processes driving galaxy evolution.

Different mechanisms, however, can produce overlapping morphological signatures. Gravitational interactions and mergers perturb both the stellar and gaseous components, producing features such as tidal tails, bridges, shells, double nuclei, and enhanced asymmetries \citep[e.g.,][]{1972ApJ...178..623T,1991ApJ...370L..65B,2004AJ....128..163L}. Their appearance depends strongly on interaction stage, mass ratio, gas content, orbital geometry, viewing angle, and surface-brightness depth. Ram-pressure stripping acts primarily on the interstellar medium as a galaxy moves through the intra-cluster or intra-group medium \citep{1972ApJ...176....1G,2006PASP..118..517B}. It can produce one-sided tails, extraplanar star-forming knots, leading-edge compression, and truncated or asymmetric star-forming discs, while leaving the old stellar component comparatively regular \citep{2016AJ....151...78P,2016MNRAS.455.2994M,2019A&A...623A..52B,2021PASA...38...35C}.

Spatially resolved observations of optically selected jellyfish candidates have confirmed that broad-band morphology can efficiently identify ram-pressure-stripping systems, particularly in cluster environments where the physical interpretation is supported by environmental information \citep{2017ApJ...844...48P,2017ApJ...844...49B,2018MNRAS.476.4753J,2019MNRAS.482.4466P}. Outside such environments, the interpretation is less straightforward. One-sided tails, asymmetric star-forming regions, clumpy emission, and disturbed outer isophotes may also arise from tidal interactions, while faint stellar tidal structures may remain undetected because of projection or limited surface-brightness sensitivity. Optical morphology may therefore identify whether a galaxy is disturbed more reliably than it identifies the physical origin of that disturbance.

The need to analyse large imaging samples has motivated several complementary approaches to galaxy morphology. Visual inspection remains effective for recognising complex and low-surface-brightness features, and citizen-science projects have extended visual classifications to large datasets \citep{2008MNRAS.389.1179L,2025A&A...694A.204C}. However, visual labels remain dependent on image quality, classifier agreement, and the adopted questions. Parametric profile fitting provides compact descriptions of regular galaxy structure but is less suited to highly asymmetric, clumpy, or multi-component systems \citep[e.g.,][]{2002AJ....124..266P,2010AJ....139.2097P,2004AJ....128..163L}. Machine- and deep-learning methods scale efficiently, but their performance depends on the quality and representativeness of the training data and may degrade when transferred between surveys with different observational properties.

Non-parametric morphological indices provide an intermediate approach that is scalable, reproducible, and physically interpretable \citep{2003ApJS..147....1C,2004AJ....128..163L,2015ApJ...814...55F,rosa2018gradient,2024MNRAS.528...82K,2025MNRAS.539.2765K}. Concentration, asymmetry, smoothness, Gini, $M_{20}$, Shannon entropy, and related quantities describe complementary aspects of the projected light distribution without requiring a smooth analytic model. Their combinations have been used to separate broad populations such as discs, spheroids, and major mergers \citep{2003ApJS..147....1C,2004AJ....128..163L,2022ApJ...938L...2F}. More recently, these indices have also been used to try to define galaxies undergoing ram-pressure-stripping phenomena, the jellyfish galaxies. \citet{2016MNRAS.455.2994M} demonstrate that non-parametric morphological parameters can be used to identify ram-pressure-stripping (RPS) candidates and showed that jellyfish galaxies occupy distinct regions in the morphological parameter space. \citet{2020MNRAS.495..554R} Showed that visually identified RPS candidates preferentially occupy the disturbed regions of classical non-parametric diagnostic diagrams, such as Concentration–Asymmetry, supporting the use of morphology as a tool for identifying stripping events. Finally, \citet{2024MNRAS.528.1125K} expanded the morphological analysis using DESI Legacy Survey imaging, introducing the concept of a morphological transition zone and showing that RPS candidates and interacting galaxies exhibit substantial overlap in non-parametric parameter space, highlighting the difficulty of separating these populations using morphology alone.

Nevertheless, the values of non-parametric indices and the boundaries of diagnostic diagrams are not universal. They depend on image depth, spatial resolution, point-spread function, signal-to-noise ratio, segmentation, and preprocessing. Regions calibrated using one dataset, particularly high-resolution imaging, therefore cannot be assumed to apply directly to wide-field ground-based surveys. Moreover, no individual index uniquely determines the physical process responsible for a disturbance. A survey-specific multivariate and probabilistic calibration is consequently required.

The DESI Legacy Imaging Surveys provide a suitable dataset for such a calibration \citep{2019AJ....157..168D}. Their large sky coverage enables statistically useful samples of both common and rare morphologies, while their depth and relatively homogeneous optical imaging allow structural measurements to be obtained consistently across large galaxy samples. The DECam-based southern imaging is also well suited to future studies combining morphology with spectroscopic and environmental information from surveys such as MAGNET and 4MOST's CHANCES and 4HS \citep{2014SPIE.9147E..0MD,2023Msngr.190...31H,2023Msngr.190...46T,2026A&A...708A.141V}.

In this work, we use ten non-parametric indices to study regular and disturbed galaxies in the DESI Legacy Imaging Surveys out to $z<0.15$. We first rank all two-dimensional combinations of the indices for separating regular spirals, regular ellipticals, and perturbed systems. We then examine whether mergers and jellyfish candidates occupy distinct regions of parameter space and define probabilistic diagnostic regions calibrated specifically for Legacy/DECam imaging. Finally, we combine all ten indices using calibrated LightGBM classifiers to test whether the structure identified in the two-dimensional diagrams persists in the full multivariate space.

Our analysis addresses two related questions: how reliably broad morphological disturbance can be identified in wide-field imaging, and how much information the projected light distribution retains about the physical origin of that disturbance. In particular, we test whether the different physical effects of gravitational interactions and ram-pressure stripping are reflected in the coherence and extent of their distributions in non-parametric parameter space.

Throughout this paper, we adopt a flat $\Lambda$CDM cosmology consistent with the final \textit{Planck} results \citep{2016A&A...594A..13P}, with $H_0=67.4~{\rm km\,s^{-1}\,Mpc^{-1}}$, $\Omega_{\rm m}=0.315$, and $\Omega_{\Lambda}=0.685$. Magnitudes are reported in the AB system.

\section{Data selection}

\subsection{Parent Legacy Survey sample}

\label{sec:parent_sample}

We select galaxies from Data Release~10 of the DESI Legacy Imaging Surveys\footnote{Data products were retrieved from Legacy Surveys Data Release~10, available at \url{https://www.legacysurvey.org}} and perform the morphological analysis using the $r$-band images, since it provides nearly complete survey coverage and a favourable compromise between depth, sensitivity to low-surface-brightness structures, and consistent tracing of the rest-frame optical morphology over $0<z<0.15$. The Legacy Surveys combine a large sky footprint with a median $5\sigma$ point-source depth of approximately $r=23.5$ and relatively homogeneous image quality, making them suitable for measuring galaxy structure across a large statistical sample.

The parent catalogue follows the selection introduced by \citet{2026A&A...709A..21S} (hereafter \citetalias{2026A&A...709A..21S}), but we summarize the criteria and their motivation here. We restrict the sample to galaxies brighter than $r=18.5$, placing the targets well above the nominal survey detection limit. This conservative magnitude threshold provides sufficient signal-to-noise not only in the central regions but also in the lower-surface-brightness outskirts, which can strongly affect pixel-based quantities such as asymmetry, clumpiness, Gini, and $M_{20}$. We also require the Legacy Survey effective-radius estimate, \texttt{shape$_r$}, to be larger than $2\arcsec$. Non-parametric indices become increasingly uncertain for barely resolved sources because the galaxy is represented by fewer independent pixels and its measured structure becomes more strongly affected by the point-spread function. The size cut therefore reduces resolution-driven biases and scatter in the morphological measurements. For comparison, the typical $r$-band PSF full width at half maximum of the DECam imaging is approximately $1.3\arcsec$, so the selected galaxies are resolved over several PSF units.

Finally, we restrict the catalogue to $z<0.15$, in order to match the upper redshift limit from the catalogs we use for labeling, which are described in the next subsections. Spectroscopic redshifts are adopted when available; otherwise, we use the Legacy Survey photometric-redshift estimate \texttt{z\_phot\_median}, following \citetalias{2026A&A...709A..21S}. After applying these magnitude, angular-size, and redshift criteria, the parent catalogue contains 1,744,454 galaxies.

\subsection{Labelled reference samples}
\label{sec:labeled_sample}

Non-parametric indices provide an objective description of the light distribution, but the mapping between structural parameter space and morphological class requires external labels \citep[e.g.][]{Conselice2003, Lotz2004, Conselice2014, 2024MNRAS.528...82K}. We therefore define labelled samples of regular spirals, regular ellipticals, mergers, and jellyfish candidates. These samples are used to calibrate the diagnostic diagrams and to train and validate the LightGBM classifiers, in order to split/identify the different morphological classes. They are not intended to constitute a complete representation of the morphological distribution observed in DECaLS. After calibration, the classifiers are applied to the full parent catalogue described above in order to assign morphology-based probabilities to all galaxies. For consistency among the labelled samples, we impose the same basic limits used throughout this work, namely $z<0.15$ and $r<18.5$. The redshift cut is chosen to remain consistent with the Galaxy Zoo 1 parent selection, while the magnitude cut ensures compatibility with the regime adopted in \citetalias{2026A&A...709A..21S}.

\subsubsection{Regular ellipticals and spirals from Galaxy Zoo 1}

The first reference samples are regular spirals and regular ellipticals. In principle, these could be selected from Galaxy Zoo DECaLS \citep{2022MNRAS.509.3966W}. However, as discussed by \citetalias{2026A&A...709A..21S}, the Galaxy Zoo DECaLS distinction between ``smooth'' and ``features/disk'' is not equivalent to a clean elliptical versus spiral separation. Moreover, the debiased\footnote{The debiased Galaxy Zoo DECaLS vote fractions are corrected for redshift-dependent classification effects. For each question in the decision tree, galaxies are grouped by intrinsic luminosity and physical size, and the vote-fraction distribution at higher redshift is mapped onto that of a low-redshift reference sample ($0.02<z<0.03$). The corrected values therefore estimate the classifications the galaxies would have received if observed at approximately ($z=0.02$).} vote fractions can differ substantially from the original ones, especially for the ``smooth'' and ``disk/features'' questions. This is shown in Appendix~\ref{app:gzdecals}, Fig.~\ref{fig:gz_comparison1}. For this reason, we do not use the debiased Galaxy Zoo DECaLS values to define the regular spiral and elliptical samples.

Instead, we use Galaxy Zoo 1 classifications \citep{2008MNRAS.389.1179L}, which are more directly matched to the classical spiral--elliptical distinction \citep{1926ApJ....64..321H}. To ensure that these galaxies are morphologically regular, we combine the Galaxy Zoo 1 labels with disturbance-related vote fractions from Galaxy Zoo DECaLS. We require an original ``No disturbance'' vote fraction greater than 0.7, a ``Major disturbance'' vote fraction smaller than 0.3, and a ``Merger'' vote fraction smaller than 0.3. We do not impose a cut on the ``Minor disturbance'' vote fraction, since this category may include weak or subjective features that do not necessarily exclude a galaxy from the regular sample. With these criteria, the final reference samples contain 5,165 regular spirals and 996 regular ellipticals. Lenticular galaxies are not treated as a separate class in this work, because they are difficult to isolate robustly with the available visual labels and can overlap with both smooth early-type and disk-dominated systems. Our regular elliptical and spiral samples should therefore be interpreted as high-purity reference samples for broad regular morphologies.

\subsubsection{Merging galaxies from Galaxy Zoo DECaLS}

We define the merger reference sample using Galaxy Zoo DECaLS classifications. In this case, the merger-related question is directly connected to the presence of strong gravitational disturbance. We use the original, non-debiased vote fractions for consistency with the other sample definitions. The comparison between original \emph{vs.} debiased fraction is presented in Fig.~\ref{fig:gz_comparison2}. We find that the difference between original and debiased merger vote fractions is smaller than for the main morphology questions. Thus, we select mergers by requiring a merger vote fraction greater than 0.7 and a ``No disturbance'' vote fraction smaller than 0.3. These cuts favour purity over completeness, retaining systems with strong visual evidence of merging activity. The final merger reference sample contains 240 galaxies. 

\subsubsection{Ram-pressure-stripping candidates from \textit{Fishing for Jellyfish Galaxies}}

\begin{figure*}
    \centering
    \includegraphics[width=0.8\textwidth]{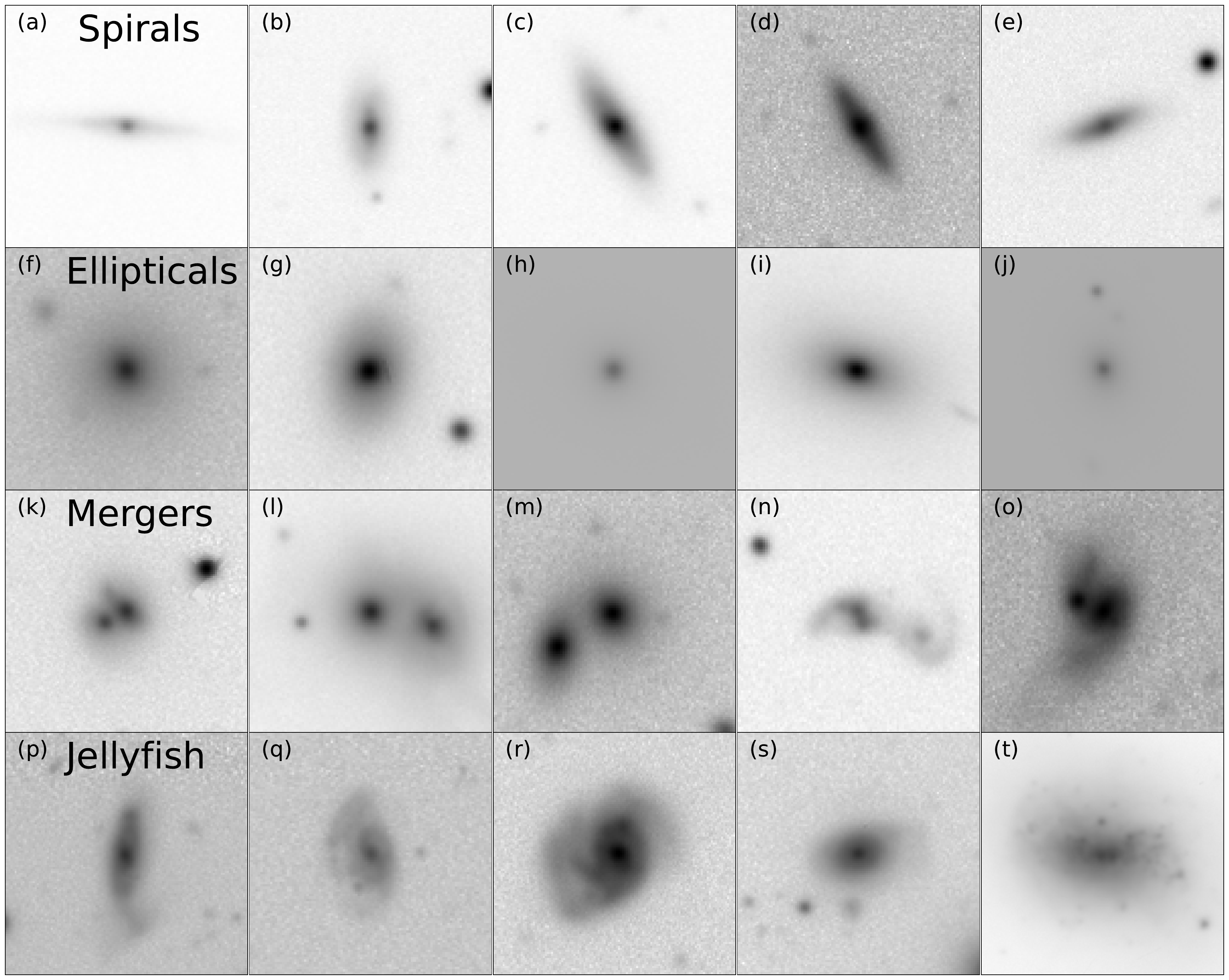}
    \caption{Galaxy images for the four morphological classes according to our labeled samples: spirals (a--e), ellipticals (f--j), mergers (k--o), and jellyfish galaxies (p--t). The images are from the r-band filter. These represent zoomed-in version of the original cutouts, which can be seen in Fig.~\ref{fig:pre_mosaic}}
    \label{fig:mosaic}
\end{figure*}

The jellyfish reference sample is selected from the citizen-science project \textit{Fishing for Jellyfish Galaxies}\footnote{Available \href{https://www.zooniverse.org/projects/cbellhouse/fishing-for-jellyfish-galaxies}{here}} \textit{(Bellhouse et al. accepted for publication)}. For the Legacy Survey data, this project was designed to identify ram-pressure-stripping candidates in DECaLS imaging. The inspected sample consists of 10,000 DECaLS DR9 images \citep{2021AAS...23723503S}, selected from bright spiral galaxies located close in projection to galaxy-cluster centres, after removing low-quality images and galaxies too small for reliable visual inspection. This selection implies that our Jellyfish sample lack red jellyfish galaxies.

We define the jellyfish sample by selecting galaxies with merger vote fraction smaller than 0.3 and either tail vote fraction greater than 0.7 or disturbed vote fraction greater than 0.7. These criteria are intended to retain galaxies with visual signatures consistent with ram-pressure stripping, while reducing contamination from systems whose disturbed appearance is more likely caused by mergers or tidal interactions. Our criteria is also supported by previous works highlighting: (i) citizen science can be used to recover reliable candidates for the ram-pressure stripping phenomena \citep{2025A&A...694A.204C}; and (ii) combination of integral field spectroscopy and optically selected jellyfish galaxies shows that $\sim 86\%$ of visually identified jellyfish galaxies are indeed undergoing RPS phenomena \citep{2025A&A...699A.357P}. 

When spectroscopic redshifts are unavailable, we use \texttt{z\_phot\_median}. With these criteria, the final jellyfish reference sample contains 932 galaxies. Figure~\ref{fig:mosaic} shows examples of galaxies from the four labelled classes. A full example of the \texttt{galmex} preprocessing steps is shown in Fig.~F1 of \citetalias{2026A&A...709A..21S}, and in Appendix~\ref{app:preproc_examples} (Fig.~\ref{fig:pre_mosaic}) we show examples of the pre-processing procedure for each morphological class. 

\subsection{The \texttt{galmex} package to estimate non-parametric indices}

The structural measurements used in this work were obtained with the \href{https://galmex.readthedocs.io/en/latest/}{\texttt{Galaxy Morphology Extractor} (\texttt{galmex})} package, following the methodology introduced and validated by \citetalias{2026A&A...709A..21S}. \texttt{galmex} is a modular Python package developed to estimate non-parametric morphological quantities in imaging surveys. It allows the main steps of the analysis, including preprocessing, segmentation, measurement, and output generation, to be handled independently.

Since non-parametric indices depend strongly on the adopted image treatment, we briefly summarize the main preprocessing steps. For each galaxy, an image cutout is created with a total length equal to six times the effective radius\footnote{The effective radius, in arcsec, is retrieved from the Legacy database, in the column \texttt{shape\_r}.} the source. The sky background is estimated from the borders of the cutout using sigma clipping, and then subtracted from the image. Source detection is performed with \texttt{SEP} \citep{2016zndo....159035B}. In this step, the central target is identified, and its centroid coordinates and basic shape parameters—namely, the semi-major and semi-minor axes and position angle—are measured. These centroid and shape measurements are used in the subsequent analysis. Neighbouring sources and foreground contaminants are removed through an isophotal interpolation procedure. Petrosian profiles and characteristic radii are then measured from the cleaned image, and the segmentation mask used for the computation of the indices is defined relative to the Petrosian radius. Namely, we only select pixels above an intensity threshold, defined as the average intensity at $k\times R_{\rm P}$. As in \citetalias{2026A&A...709A..21S}, we adopt a conservative threshold of $k=1$ for the segmentation mask definition, which corresponds roughly to a $5\sigma$ threshold above the background (see Fig.~\ref{fig:kvariation_thresh} in Appendix~\ref{app:metrics_variation}).

A difference with respect to \citetalias{2026A&A...709A..21S} is that here we expand the set of asymmetry- and smoothness-related quantities in order to better describe disturbed morphologies. In addition to the classical CAS asymmetry, hereafter $A_1$ \citep{2003ApJS..147....1C}, and the shape asymmetry, $A_3$ \citep{2016MNRAS.456.3032P}, we also include a correlation-based asymmetry, $A_2$, following \citet{2015ApJ...814...55F}. This quantity is computed as $1-r$, where $r$ is the Spearman correlation coefficient \citep{Spearman1904} between the original pixel values and those of the image rotated by $180^\circ$. We adopt an analogous definition for the correlation-based smoothness, $S_2$, also following \citet{2015ApJ...814...55F}, where the correlation is measured between the original and smoothed images. The ten indices used in this work are:

\begin{enumerate}
    \item \textbf{Concentration ($C$)} -- Measures how centrally concentrated the galaxy light distribution is \citep{2003ApJS..147....1C}.

    \item \textbf{Pixel-wise asymmetry ($A1$)} -- Quantifies rotational asymmetry by comparing the original image with a version rotated by $180^\circ$ \citep{2003ApJS..147....1C}.

    \item \textbf{Correlation-based asymmetry ($A2$)} -- Measures asymmetry from the correlation between the original and $180^\circ$-rotated images \citep{2015ApJ...814...55F, 2020A&C....3000334B}.

    \item \textbf{Shape asymmetry ($A3$)} -- Measures asymmetry using the binary detection mask, making it sensitive to low-surface-brightness asymmetric features \citep{2016MNRAS.456.3032P}.

    \item \textbf{Pixel-wise smoothness ($S1$)} -- Measures small-scale structure by comparing the original image with a smoothed version \citep{2003ApJS..147....1C}.

    \item \textbf{Correlation-based smoothness ($S2$)} -- Quantifies small-scale structure from the correlation between the original and smoothed images \citep{2015ApJ...814...55F, 2020A&C....3000334B}.

    \item \textbf{Second-order moment of light ($M{20}$)} -- Measures how the brightest 20\% of the galaxy light is spatially distributed relative to the total second-order moment \citep{2004AJ....128..163L, 2025MNRAS.539.2765K}.

    \item \textbf{Shannon entropy ($E$)} -- Measures the complexity of the light distribution from the spread of pixel intensities \citep{2015ApJ...814...55F, 2020A&C....3000334B, 2024MNRAS.528...82K}.

    \item \textbf{Gini index ($Gini$)} -- Quantifies how unevenly the total galaxy flux is distributed among the pixels \citep{2004AJ....128..163L, 2024MNRAS.528...82K}.

    \item \textbf{Gradient pattern asymmetry ($G2$)} -- Measures asymmetries in the local gradient field of the image \citep{rosa2018gradient, 2020A&C....3000334B, 2024MNRAS.528...82K}.
\end{enumerate}

Hereafter, we refer to the combination $C+A1+A2+A3+S1+S2$ as the extended CAS system, whereas $M{20}+E+Gini+G2$ defines the MEGG system. Since these metrics are expected to depend on the segmentation mask adopted, we also show in Appendix~\ref{app:metrics_variation} the metrics variation (Figs.~\ref{fig:kvariation_cas} and \ref{fig:kvariation_megg}) as a function of adopted $k$ in the segmentation process.

\section{Metrics distribution}

\begin{figure*}
    \centering
    \includegraphics[width=0.95\textwidth]{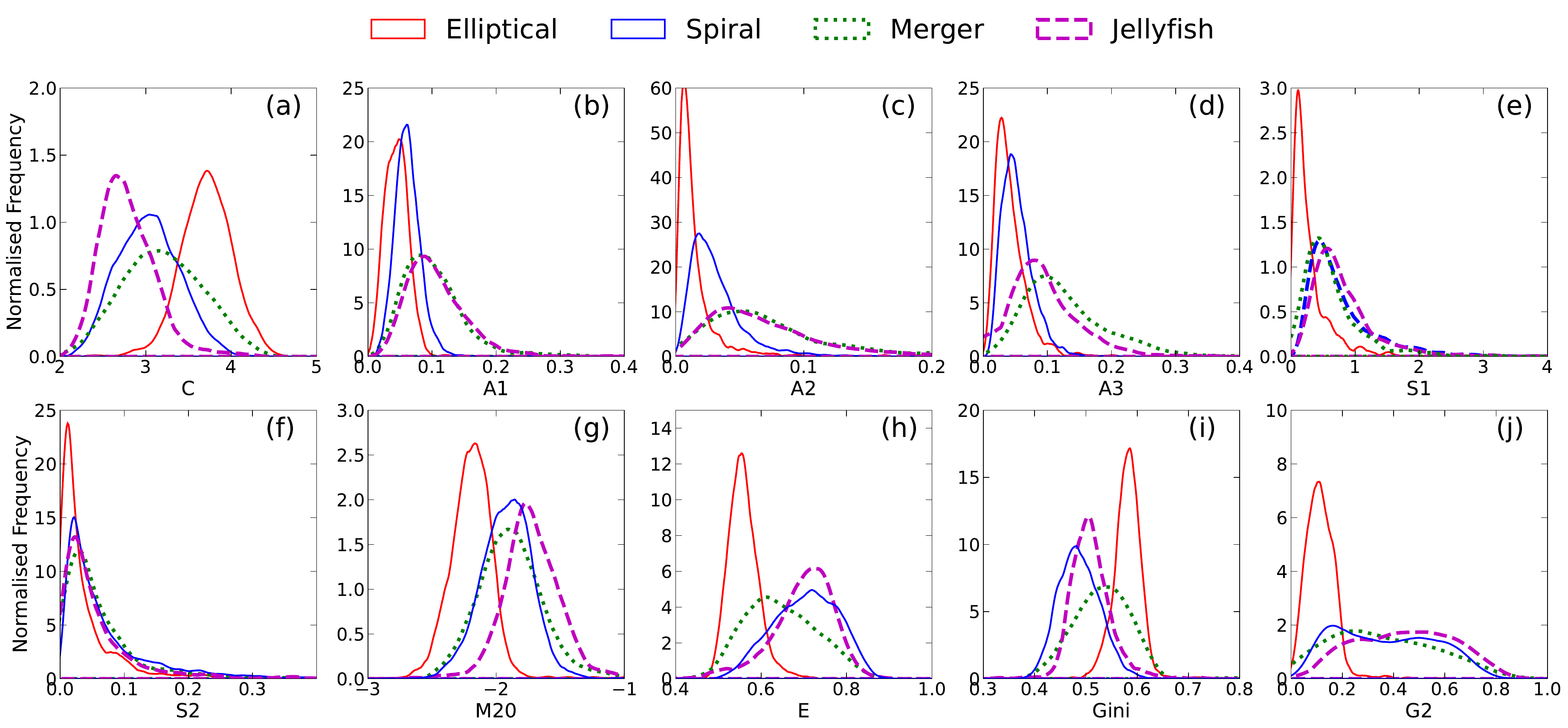}
    \caption{Normalised distributions of the ten non-parametric morphological indices for regular ellipticals (solid red), regular spirals (solid blue), mergers (dotted green), and jellyfish galaxies (dashed magenta). The panels show concentration $C$ (a), pixel-wise asymmetry $A1$ (b), correlation-based asymmetry $A2$ (c), shape asymmetry $A3$ (d), pixel-wise smoothness $S1$ (e), correlation-based smoothness $S2$ (f), $M_{20}$ (g), Shannon entropy $E$ (h), Gini index $G$ (i), and gradient pattern asymmetry $G2$ (j).}
    \label{fig:metrics_dist}
\end{figure*}

Figure~\ref{fig:metrics_dist} shows the normalised distributions of the ten non-parametric indices for the four reference samples. Regular ellipticals occupy the highest-concentration regime, whereas spirals extend towards lower $C$. The asymmetry and smoothness indices, $A1$, $A2$, $A3$, $S1$, and $S2$, show considerable overlap between the two regular classes. Concentration and the MEGG indices therefore provide the clearest individual separation between regular spirals and ellipticals, consistent with the trends reported by \citetalias{2026A&A...709A..21S}.

Mergers and jellyfish galaxies extend towards higher asymmetry and smoothness than the regular populations, but overlap strongly with one another in these quantities. Jellyfish candidates show, on average, lower concentration, less negative $M{20}$, higher entropy, lower Gini values, and slightly higher $G2$ than mergers. Mergers instead span a broader range and more frequently occupy intermediate values between the regular spiral and elliptical distributions. These trends motivate the multivariate diagnostic tests presented in the following sections.

\section{Using non-parametric metrics to separate morphological classes}

Two-dimensional diagrams provide an interpretable way to examine which properties of the light distribution drive the separation between morphological classes. We therefore rank all pairs of non-parametric indices before combining the full set of measurements in Sect.~5.

\subsection{Selection of optimal two-dimensional diagnostic diagrams}

To identify the most informative two-dimensional diagnostic diagrams, we evaluated all possible pairs of non-parametric indices in a three-class classification problem, where the classes are spirals, ellipticals, and perturbed systems, the latter defined as the combination of mergers and jellyfish galaxies. For each parameter pair \((x,y)\), the classification performance was then estimated through stratified \(K\)-fold cross-validation, with \(K=5\). In this procedure, the full sample is divided into \(K\) disjoint subsets, or \emph{folds}, each containing approximately the same class proportions as the original data set. At each iteration, one fold is kept aside as an independent test sample, while the remaining \(K-1\) folds are used as the training sample. This process is repeated \(K\) times, so that each galaxy is classified exactly once in a test fold and \(K-1\) times contributes to the training set. In this way, the performance of each 2D diagram is evaluated on statistically independent data, reducing the sensitivity to any particular train--test split.

Within each training set, we modeled the distribution of each class \(c\) in the \((x,y)\) plane using a two-dimensional kernel density estimator,
\begin{equation}
    \hat{f}_c(\mathbf{z}) = \frac{1}{N_c}\sum_{i=1}^{N_c} K_h(\mathbf{z}-\mathbf{z}_i),
\end{equation}
where \(\mathbf{z}=(x,y)\) is the point in parameter space where the density is evaluated, \(\mathbf{z}_i=(x_i,y_i)\) is the position of the \(i\)-th training galaxy in class \(c\), \(N_c\) is the number of training galaxies in that class, and \(K_h\) is a kernel of bandwidth \(h\). The estimated density at \(\mathbf{z}\) is therefore obtained by summing the contributions of all training galaxies in class \(c\), weighted according to their distance from that point. Before fitting the KDEs, each parameter was standardised using the mean and standard deviation of the training fold. The same transformation was then applied to the corresponding test fold. This ensures that the KDE bandwidth has the same meaning for all two-dimensional projections and prevents parameters with larger numerical ranges from dominating the density estimate. We adopted a Gaussian kernel with fixed bandwidth \(h=0.10\), chosen after testing values in the range \(0.10 \leq h \leq 0.50\), for which we find no significant impact in our results (see Appendix \ref{app:smoothing}).

The posterior probability of class membership was then computed as
\begin{equation}
    P(c \mid \mathbf{z}) \propto \pi_c\,\hat{f}_c(\mathbf{z}),
\end{equation}
where \(\pi_c\) is the prior probability assigned to class \(c\). In this work we adopted equal priors, \(\pi_c=1/3\), for all classes, implying that the prior have a normalization role, rather to attribute different weights to different classes. This ensures that the inferred posterior probabilities are driven mainly\footnote{Small variations can still occur due to the smoothing factor adopted.} by the morphology-dependent distribution of galaxies in the \((x,y)\) plane, rather than by the unequal sizes of the different subsamples.
Each galaxy in the test fold was then assigned to the class that maximizes this posterior probability,
\begin{equation}
    \hat{c}(\mathbf{z}) = \arg\max \left[\pi_c\,\hat{f}_c(\mathbf{z})\right].
\end{equation}
After repeating this procedure across the five folds, the predictions were combined into a single confusion matrix for each parameter pair. From this matrix, we computed the recall, defined as the fraction of true objects in a given class that were correctly identified, and the balanced accuracy, defined as the average recall across all classes. We adopted the balanced accuracy (BA) as our main ranking statistic,
\begin{equation}
    \mathrm{BA} = \frac{1}{3}\sum_{c=1}^{3} \frac{TP_c}{TP_c+FN_c},
\end{equation}
where \(TP_c\) and \(FN_c\) are the true positives and false negatives for class \(c\), respectively. This metric corresponds to the mean recall over the three classes and is therefore particularly well suited to our problem, as it gives equal weight to spirals, ellipticals, and perturbed galaxies, independently of their relative abundances. 

We ranked the candidate diagrams according to their balanced accuracy (BA), which corresponds to the mean class-wise recall and therefore quantifies how well the three morphological classes are recovered on average, while mitigating the effect of class imbalance. In addition to the median balanced accuracy, we also report the median recall of spirals, ellipticals, and perturbed galaxies, and the median Macro-F1, defined as
\begin{equation}
\mathrm{Macro\text{-}F1} = \frac{1}{3}\left(F1_{\rm S} + F1_{\rm E} + F1_{\rm P}\right),
\end{equation}
where the class-wise F1 score is defined as
\begin{equation}
F1_i = \frac{2\,P_i\,R_i}{P_i + R_i},
\end{equation}
with $P_i$ and $R_i$ denoting the precision and recall of class $i$. Notably, these results can depend on the thresholds adopted in sample selection, particularly for the perturbed (mergers + jellyfish systems) subsample. We therefore repeat this analysis varying threshold adopted for subsample selection, but we find no significant change in the performance of the 2-D diagrams, with a maximum variation of 4\% in the balanced accuracy. It is also important to highlight that imposing too strict selection criteria strongly affects the sample size. For instance, if we adopt 0.9 as a threshold in our merger selection, we would be left with a sample of 63 merging galaxies, which prevents the statistical analysis presented in this paper. 

\begin{table}
    \centering
    \caption{Top-performing 2D parameter combinations ranked according to the balanced accuracy. Columns respectively show: (1) metric used as x coordinate; (2) metric used as y coordinate; (3) the balanced accuracy (BA); (4) the Macro-F1 score; (5) the recall for spirals ($\rm R_S$); (6) the recall for ellipticals ($\rm R_E$); and (7) the recall for perturbed galaxies ($\rm R_P$).}
    \label{tab:best_2d_metrics_ba}
    \begin{tabular}{l l c c c c c}
        \hline
        $x$ & $y$ & BA & Macro-F1 & $\rm R_S$ & $\rm R_E$ & $\rm R_P$ \\
        \hline
        $M20$ & $E$   & 0.77 & 0.67 & 0.69 & 0.94 & 0.67 \\
        $A3$  & $E$   & 0.77 & 0.70 & 0.74 & 0.93 & 0.64 \\
        $A2$  & $E$   & 0.76 & 0.69 & 0.72 & 0.93 & 0.64 \\
        $A1$  & $E$   & 0.76 & 0.68 & 0.71 & 0.91 & 0.65 \\
        $C$  & $A1$  & 0.75 & 0.67 & 0.70 & 0.88 & 0.67 \\
        $C$   & $A2$  & 0.75 & 0.66 & 0.69 & 0.90 & 0.66 \\
        $C$   & $A3$  & 0.75 & 0.66 & 0.68 & 0.87 & 0.69 \\
        $A1$  & $G2$  & 0.74 & 0.63 & 0.64 & 0.94 & 0.65 \\
        $A3$  & $G2$  & 0.73 & 0.61 & 0.60 & 0.93 & 0.66 \\
        $A2$  & $M20$ & 0.73 & 0.62 & 0.62 & 0.89 & 0.67 \\
        \hline
    \end{tabular}
\end{table}

The top-performing two-dimensional combinations are listed in Table~\ref{tab:best_2d_metrics_ba}. Shannon entropy, $E$, appears in each of the four highest-ranked diagrams, indicating that the distribution of pixel intensities contains substantial information about broad morphological class. In the adopted segmentation masks, regular ellipticals are generally dominated by a steep and centrally organised light distribution, whereas spiral arms, star-forming regions, tidal structures, and diffuse outer emission contribute pixels over a wider range of intensities in discs and disturbed systems. These structures produce a less uniform intensity distribution across the galaxy and therefore modify the measured entropy. Yet, because the entropy depends on which low-intensity pixels are included in the segmentation mask, its absolute value and the corresponding diagnostic boundaries remain dependent on image depth and preprocessing.

Entropy alone, however, does not encode where pixels of a given intensity are located. Its strong performance in combination with $M{20}$ and the asymmetry indices can therefore be understood as the combination of complementary information. The $M{20}$ statistic describes the spatial distribution of the brightest regions, while $A1$, $A2$, and $A_3$ quantify different forms of rotational or shape asymmetry. The $E$--$M{20}$ and $E$--$A3$ planes can consequently distinguish galaxies with comparable levels of intensity complexity but different spatial organisations of their light. We therefore focus below on these two combinations, which reach balanced accuracies of approximately $0.77$.

The location of perturbed galaxies towards higher asymmetry and less negative $M{20}$ values is consistent with previous studies of ram-pressure-stripping candidates. Using HST imaging at $0.3<z<0.7$, \citet{2016MNRAS.455.2994M} showed that combinations of concentration, asymmetry, Gini, and $M{20}$ parameters can efficiently exclude regular galaxies and identify a smaller population of morphologically disturbed RPS candidates. Their analysis also showed, however, that morphology-based cuts remain substantially contaminated by galaxies disturbed for other reasons. At lower redshift, \citet{2020MNRAS.495..554R} found that visually selected stripping candidates in Coma have, on average, higher asymmetry, lower concentration and Gini, and less negative $M{20}$ than the remaining cluster population. Our results recover these qualitative trends using DECam $r$-band imaging over a much larger and more heterogeneous sample.

\begin{figure*}
    \centering
    \includegraphics[width=0.8\textwidth]{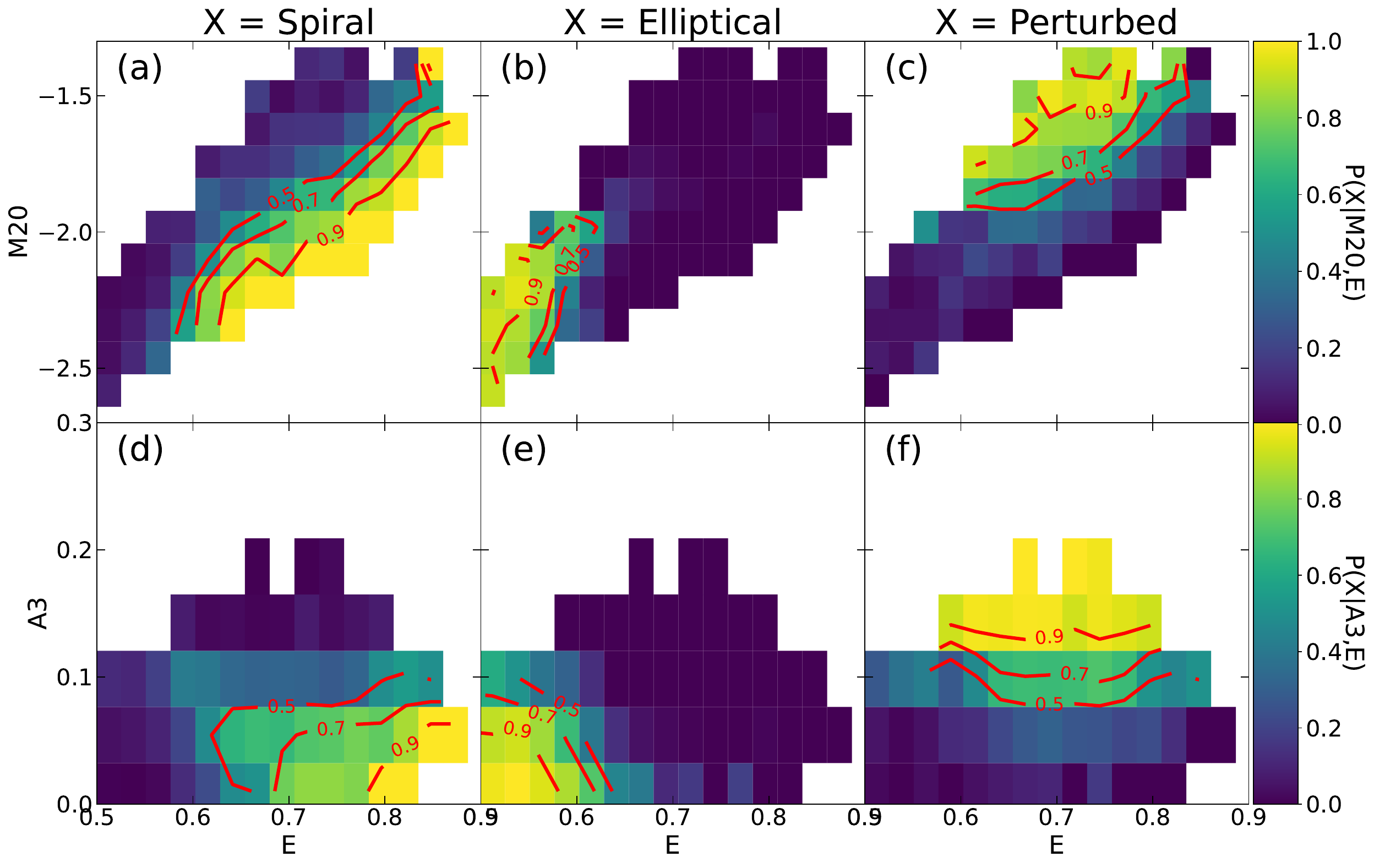}
    \caption{Posterior probability maps in the $E$--$M{20}$ plane (top panels) and in the $E$--$A3$ plane (bottom panels) for the, from left to right, spirals (panels a and d), ellipticals (panel b and e) and perturbed systems (panel c and f), respectively. Each pixel colour indicates the posterior probability, $P(X\,|\,M{20},E)$ and $P(X\,|\,A3,E)$, respectively, where $X$ denotes the morphological class. The red lines indicate the contours containing 50, 70 and 90\% probability for the respective class.}
    \label{fig:distribution_first}
\end{figure*}

To visualize how the different morphological classes populate a given two-dimensional parameter space, we constructed posterior probability grids in the selected planes, following a simple histogram-based Bayesian scheme. In Fig.~\ref{fig:distribution_first}, we show the probability distribution of each class in the 2D diagrams for the two combinations with the highest balanced accuracy, i.e. $E$ \emph{vs.} $M20$, and $E$ \emph{vs.} $A3$. Panels (a) to (f) show the pixel-wise probability of a galaxy being spiral, elliptical, and perturbed, respectively, given a pair of non-parametric indexes values ($E$ \emph{vs.} $M20$ on top, and $E$ \emph{vs.} $A3$ in the bottom). In the following, we describe the procedure to define the probability in each pixel. For a given pair of non-parametric indices, \((x,y)\), we first built a common rectangular grid with \(N_{\mathrm{bin}}\) cells covering the full range spanned by the spiral, elliptical, and perturbed subsamples. For each class \(c \in \{\mathrm{Sp}, \mathrm{Ell}, \mathrm{Per}\}\), we then computed the two-dimensional histogram \(N_{ij}^{(c)}\), where \(i\) and \(j\) label the bin in the \(x\)- and \(y\)-directions, respectively. The class-conditional probability in each cell was estimated as
\begin{equation}
    \hat{p}(\mathbf{z}_{ij}\mid c) =
    \frac{N_{ij}^{(c)} + \epsilon}
         {N^{(c)} + \epsilon\,N_{\mathrm{bin}}},
\end{equation}
where \(\mathbf{z}_{ij}=(x_i,y_j)\), \(N^{(c)}=\sum_{ij}N_{ij}^{(c)}\) is the total number of galaxies in class \(c\), and \(\epsilon\) is a small smoothing term introduced to avoid zero-probability bins, and set to $10^{-6}$. In practice, this corresponds to a regularized empirical estimate of the probability density for each class in the chosen 2D plane. To avoid noisy estimates in sparsely populated regions, we masked all cells containing fewer than \(N_{\mathrm{min}}=10\) galaxies in the combined sample. The resulting grids therefore provide, for each location in the \((x,y)\) plane, the probability that a galaxy belongs to the spiral, elliptical, or perturbed class. 

\begin{figure}
    \centering
    \includegraphics[width=0.7\linewidth]{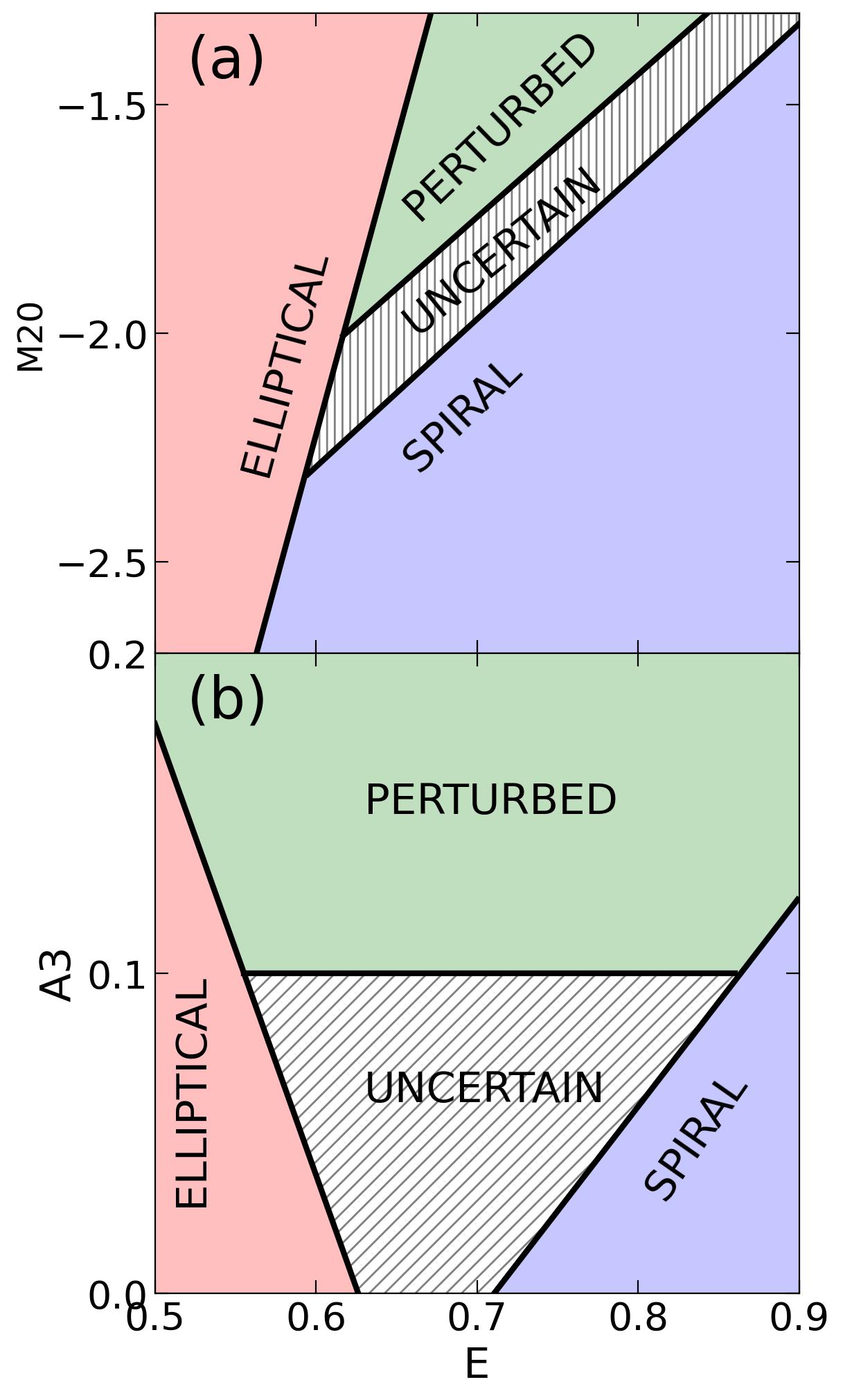}
    \caption{Proposed regions in the $M20$ \emph{vs} $E$ (panel a) and $A3$ \emph{vs} $E$ (panel b) to separate between regular ellipticals, spirals and perturbed systems. In comparison to previous works, we also add an extra region for "uncertain" galaxies. The equations for each line can be checked in the text.}
    \label{fig:proposed_diagram_first}
\end{figure}

In the same figure, we show the contours representing the 50, 70 and 90\% probabilities for each different class. Particularly, we use the 70\% line to define regions in these diagrams that can be used by the astronomical community to separate mainly regular spirals and ellipticals from perturbed systems. We highlight that this is the first time that this kind of diagram is proposed with a probabilistic approach, whereas the diagrams shown in, for example, \citet{Conselice2003} and \citet{Lotz2004} are based on visual inspection. Besides, we also allow for an "uncertain" region, in order to enhance purity and avoid forcing classifications in regions with mixed populations. Our proposed diagram is shown in Fig.~\ref{fig:proposed_diagram_first}, in which we use the 70\% contours as reference to define the separating lines. The lines equations are given below:

\begin{itemize}
    \item $M20$ \emph{vs} $E$: 
    \begin{enumerate}
        \item $M20 = 3.12\,E - 3.93$, valid for $0.617 \leq E \leq 0.870$.
        \item $M20 = 3.23\,E - 4.23$, valid for $0.595 \leq E \leq 0.900$.
        \item $M20 = 12.93\,E - 9.98$, valid for $0.500 \leq E \leq 0.850$.
    \end{enumerate}
    \item $A3$ \emph{vs} $E$:
    \begin{enumerate}
        \item $A3 = -1.41\,E + 0.883$, valid for $0.500 \leq E \leq 0.900$.
        \item $A3 = 0.653\,E - 0.464$, valid for $0.500 \leq E \leq 0.900$.
        \item $A3 = 0.1$, valid for $0.555 \leq E \leq 0.860$.
    \end{enumerate}
    
\end{itemize}

\subsection{Separating mergers and jellyfish galaxies}

Having established that non-parametric morphology can separate regular from perturbed systems, we now ask whether the structure of the optical light also retains information about the physical origin of the disturbance. This comparison is not only a classification exercise. Mergers and ram-pressure stripping act through different physical mechanisms and over different timescales, so the extent to which each population occupies a compact or broad region of parameter space is itself astrophysically informative. We therefore compare mergers and jellyfish candidates first as a two-class problem and then examine how both populations overlap with regular spirals and ellipticals.

\begin{figure*}
    \centering
    \includegraphics[width=0.9\textwidth]{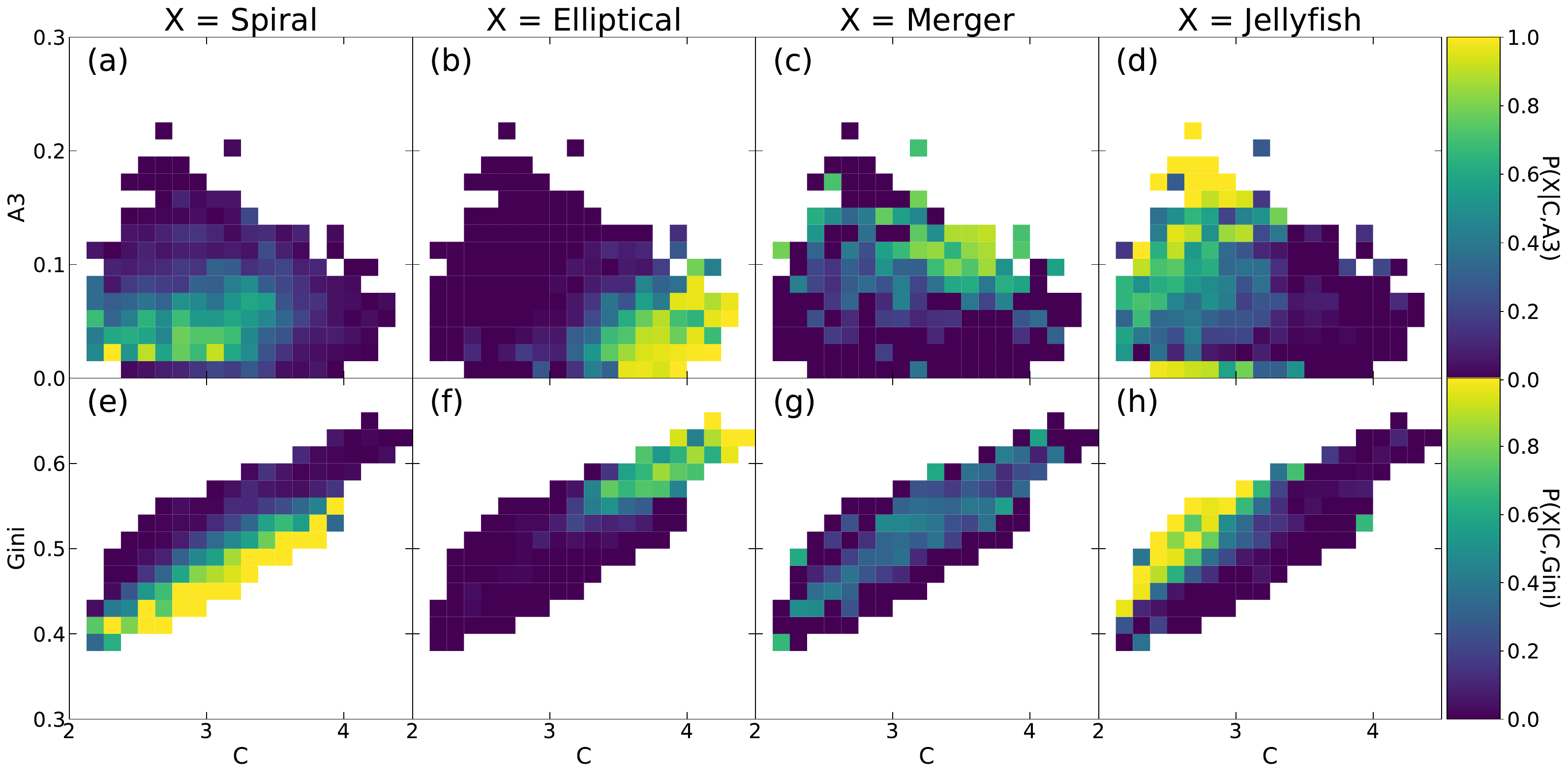}
    \caption{Posterior probability maps for regular spirals, regular ellipticals, mergers, and jellyfish galaxies in the $C$--$A3$ plane (top panels) and the $C$--$Gini$ plane (bottom panels). Jellyfish candidates occupy a comparatively coherent low-concentration region, whereas merger probabilities are distributed over a broader area and overlap substantially with the other morphological classes. Because the merger boundaries are highly non-linear and discontinuous, we do not draw a single set of four-class probability contours in the $C$--$A3$ plane.}
    \label{fig:distribution_second}
\end{figure*}

We repeat the analysis for the two-class problem of separating mergers from jellyfish candidates. The best-performing combinations are listed in Table~\ref{tab:best_2d_metrics_merger_jellyfish}. Concentration, $C$, appears in all of the highest-ranked pairs, showing that the radial distribution of the optical light provides the principal distinction between the two disturbed samples. The $C$--$A3$ and $C$--Gini planes both reach a balanced accuracy of $0.73$. The former recovers jellyfish galaxies with $R_{\rm JF}=0.80$ and mergers with $R_{\rm M}=0.66$, whereas the latter gives $R_{\rm JF}=0.77$ and $R_{\rm M}=0.68$.

More generally, jellyfish recall exceeds merger recall in all of the top-ranked combinations: the listed diagrams recover approximately $77$--$83\%$ of the jellyfish candidates, compared with approximately $62$--$68\%$ of the mergers. This systematic difference is not specific to one choice of indices. Instead, it indicates that the jellyfish candidates occupy a more coherent region of non-parametric parameter space, whereas mergers are distributed over a wider range of structural configurations. The difference in separability is therefore itself an astrophysical result, rather than only a measure of classifier performance.

The importance of concentration is physically plausible because mergers and optically selected jellyfish galaxies redistribute visible light in different ways. Merging systems may contain two overlapping stellar components, prominent bulges, or multiple bright nuclei, while merger-driven gas inflows can enhance centrally concentrated star formation. These effects can preserve or increase the fraction of the total light located near the centre. Ram-pressure stripping acts primarily on the gas rather than directly on the old stellar component, but during the optically prominent jellyfish phase it can produce luminous off-centre star-forming regions, asymmetric outer-disc emission, and extended tails. In the $r$ band, these structures add light at large radii and can therefore reduce the measured concentration.

The greater coherence of the jellyfish population may reflect the comparatively directional nature of ram-pressure stripping. Although the strength and visibility of stripping depend on orbital geometry, gas content, and viewing angle, the process preferentially displaces gas and recent star formation towards one side of the galaxy while leaving much of the pre-existing stellar disc comparatively regular. The combination of a recognisable disc with asymmetric, extended, and off-centre emission can therefore produce a relatively consistent non-parametric signature. 

\begin{table}
    \centering
    \caption{Top-performing 2D parameter combinations for the merger--jellyfish separation, ranked according to the balanced accuracy. Columns respectively show: (1) metric used as x coordinate; (2) metric used as y coordinate; (3) the balanced accuracy (BA); (4) the Macro-F1 score; (5) the recall for mergers ($\rm R_M$); and (6) the recall for jellyfish galaxies ($\rm R_J$).}
    \label{tab:best_2d_metrics_merger_jellyfish}
    \begin{tabular}{l l c c c c}
        \hline
        $x$ & $y$ & BA & Macro-F1 & $\rm R_M$ & $\rm R_J$ \\
        \hline
        $C$   & $A3$   & 0.73 & 0.68 & 0.66 & 0.80 \\
        $C$   & $Gini$ & 0.73 & 0.66 & 0.68 & 0.77 \\
        $C$   & $A2$   & 0.73 & 0.69 & 0.63 & 0.83 \\
        $C$   & $S1$   & 0.72 & 0.66 & 0.67 & 0.77 \\
        $C$   & $M20$  & 0.72 & 0.66 & 0.67 & 0.77 \\
        $C$   & $S2$   & 0.72 & 0.67 & 0.65 & 0.79 \\
        $C$   & $A1$   & 0.72 & 0.68 & 0.62 & 0.81 \\
        $S2$  & $E$    & 0.71 & 0.65 & 0.64 & 0.77 \\
        $C$   & $E$    & 0.70 & 0.65 & 0.63 & 0.78 \\
        $A2$  & $E$    & 0.70 & 0.65 & 0.63 & 0.78 \\
        \hline
    \end{tabular}
\end{table}

Figure~\ref{fig:distribution_second} reveals a clear difference in the morphology of the two disturbed populations. Jellyfish candidates preferentially occupy a comparatively coherent region extending towards low $C$, high $A3$, and lower $Gini$ values. Mergers, by contrast, span most of the populated parameter space and overlap with regular spirals, ellipticals, and jellyfish galaxies. The secondary index changes the relative recovery of the two disturbed classes, but does not alter this fundamental contrast.

The comparatively coherent jellyfish distribution is one of the main astrophysical results of this analysis. It suggests that ram-pressure stripping, at least during the optically sample selected here, produces a more concise imprint on the projected light distribution than gravitational interactions. The simultaneous presence of a recognizable stellar disc, reduced concentration, and asymmetric in the outskirts places many jellyfish candidates in a similar part of non-parametric parameter space. Their higher recovery is therefore a consequence of this morphological coherence, rather than simply an advantage of a particular diagnostic diagram.

The broad merger distribution is also a physical result rather than merely a limitation of the method. A merger is not a single morphological state, but an evolutionary sequence whose observable structure changes substantially with interaction stage. Before or near first passage, the system may retain two largely regular stellar components. Strong tidal features and displaced bright regions can subsequently increase asymmetry and produce less negative $M{20}$ values. During coalescence, overlapping nuclei and centrally concentrated star formation may instead increase concentration and Gini. At later stages, tidal features can fade and the remnant may again appear comparatively smooth and concentrated.

In addition to merger stage, the measured morphology depends on mass ratio, gas fraction, orbital geometry, viewing angle, and the surface-brightness depth at which tidal structures can be detected. The visually selected merger sample is physically closer to a sample of strongly interacting systems and is not restricted to a single stage or interaction strength. It is therefore expected to include disc-like, spheroid-like, double-nucleus, and highly irregular systems. The absence of a compact merger locus reflects this genuine astrophysical diversity and explains why projected morphology alone cannot isolate mergers with the same reliability as jellyfish candidates.

The overlap between stripping and gravitationally disturbed systems is consistent with the Legacy-based analysis of \citet{2024MNRAS.528.1125K}. They found that the strongest RPS candidates, particularly those with JClass $\geq 3$, separate from isolated galaxies in diagrams involving asymmetry, concentration, and S\'ersic index. Their interacting galaxies, however, occupied a broad region extending across both the isolated sequence and the proposed morphological transition zone. Their diagnostic diagrams were therefore effective for identifying morphologically disturbed RPS candidates, but were not sufficient by themselves to distinguish ram-pressure stripping from tidal interactions.

\begin{figure}
    \centering
    \includegraphics[width=0.7\linewidth]
    {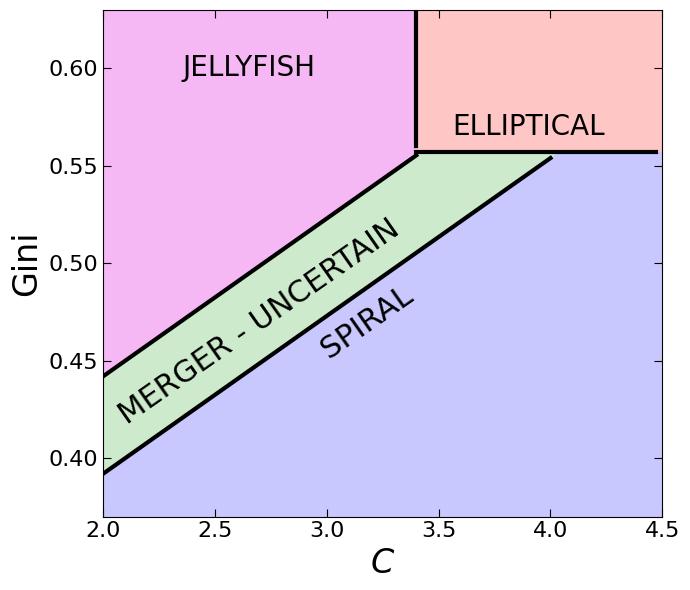}
    \caption{Empirical regions in the $C$--$Gini$ plane for regular spirals, regular ellipticals, mergers, and jellyfish candidates. Regular galaxies and jellyfish candidates occupy comparatively coherent regions, whereas mergers remain distributed across overlapping parts of parameter space. The merger-labelled region should therefore be interpreted as a region of enhanced probability rather than as a clean merger selection.}
    \label{fig:proposed_diagram_second}
\end{figure}

Our analysis extends this result by comparing jellyfish candidates directly with a broader visually selected merger population and by quantifying the class distributions probabilistically. We find that the overlap is asymmetric: jellyfish candidates occupy a comparatively coherent region of non-parametric parameter space, whereas mergers extend across regions associated with several regular and disturbed morphologies. This contrast persists in different two-dimensional projections and is therefore not specific to a single empirical boundary. The different degrees of morphological separation are also reflected in the empirical diagnostic regions. In the $C$--$Gini$ plane, regular spirals, regular ellipticals, and jellyfish candidates occupy sufficiently coherent regions to define useful boundaries, whereas mergers remain distributed across several overlapping parts of the diagram. Figure~\ref{fig:proposed_diagram_second} presents the resulting four-class diagnostic diagram. The adopted boundaries are:

\begin{enumerate}
    \item $\mathrm{Gini} = 0.081C + 0.23$, for
    $2.15 \leq C \leq 3.98$;
    \item $C = 3.40$, for
    $0.56 \leq \mathrm{Gini} \leq 0.63$;
    \item $\mathrm{Gini} = 0.557$, for
    $3.40 \leq C \leq 4.47$;
    \item $\mathrm{Gini} = 0.081C + 0.28$, for
    $2.22 \leq C \leq 3.40$.
\end{enumerate}

The merger-labelled region should therefore be interpreted as a region of enhanced merger probability rather than as a complete or pure merger selection. Conversely, the more compact jellyfish region reflects the greater morphological coherence of the optically selected stripping candidates.

\section{Combining non-parametric indices into calibrated class probabilities}
\label{sec:lightgbm}

The two-dimensional diagrams in Sect.~4 provide an interpretable view of the structure of the morphological parameter space. They show that regular and perturbed galaxies can be separated using a small number of complementary indices, while also revealing an important difference within the disturbed population: jellyfish candidates occupy a comparatively coherent region, whereas mergers span a much broader range of structural configurations. Each diagram, however, represents only one projection of the full ten-dimensional parameter space.

We therefore complement the diagnostic diagrams with calibrated LightGBM classifiers that combine all ten non-parametric indices simultaneously. This allows us to address two related questions: how much the classification improves when the complete morphological information is used, and whether the contrasting parameter-space distributions of jellyfish galaxies and mergers persist in the full multivariate space. The resulting probabilities should be interpreted as morphology-based probabilities conditional on the labelled reference samples and the observational properties of the Legacy imaging.

\subsection{Regular ellipticals, regular spirals, and perturbed galaxies}

We first consider a three-class experiment separating regular ellipticals, regular spirals, and a combined perturbed population composed of mergers and jellyfish galaxies. This extends the two-class LightGBM analysis presented in \citetalias{2026A&A...709A..21S}. We retain spirals and ellipticals as separate classes because regular discs and spheroids define distinct structural reference populations. Perturbed galaxies can therefore be identified according to how their morphologies depart from either regular sequence.

The classifier uses the full set of ten non-parametric indices $(C,A1,A2,A3,S1,S2,M{20},E,Gini,G2)$. Class imbalance is treated with SMOTE, applied only within the training portion of each fold. The raw LightGBM outputs are then calibrated using a sigmoid transformation with five-fold stratified cross-validation. All performance statistics reported below are measured on an independent test sample that was used neither for training nor for probability calibration. Further details of the model configuration and calibration procedure are given in Appendix~\ref{app:lightgbm}.

\begin{figure}
    \centering
    \includegraphics[width=\linewidth]{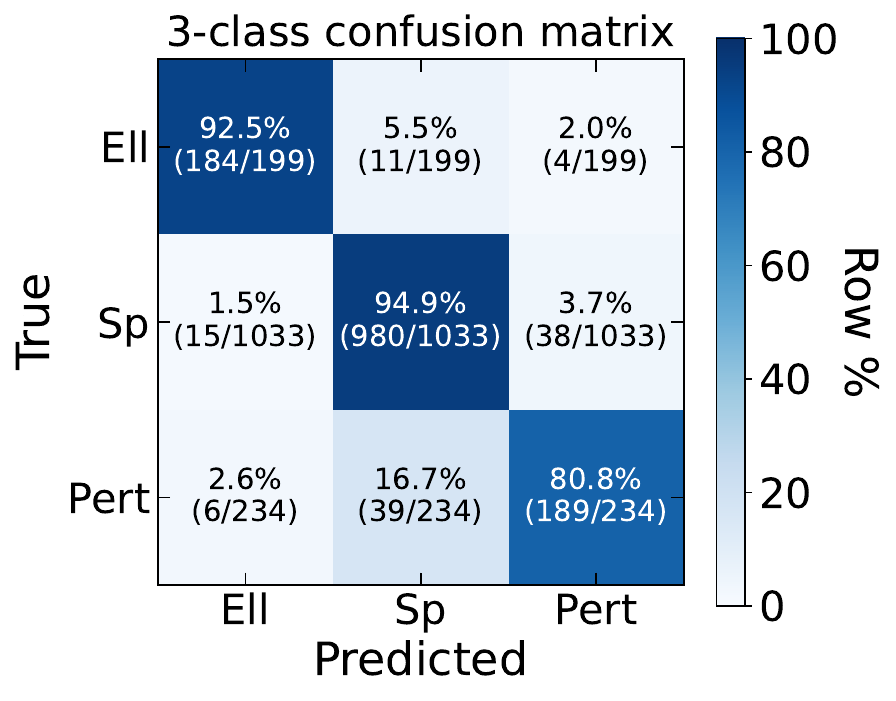}
    \caption{Confusion matrix of the calibrated three-class LightGBM classifier evaluated on the independent test sample. The model separates regular ellipticals, regular spirals, and perturbed galaxies. Each row is normalised by the number of galaxies in the corresponding true class. Percentages therefore give the fraction assigned to each predicted class, while the values in parentheses show the corresponding numbers of galaxies.}
    \label{fig:cm_3class}
\end{figure}

The performance of the calibrated three-class classifier is summarised in Table~\ref{tab:lightgbm_3class_performance} and Fig.~\ref{fig:cm_3class}. The model reaches an overall accuracy of $0.92$, a balanced accuracy of $0.89$, and a macro-averaged F1-score of $0.89$. The similarity between the overall and balanced accuracies indicates that the performance is not driven only by the numerical dominance of spirals.

Regular spirals and ellipticals are recovered with F1-scores of $0.95$ and $0.91$, respectively. The perturbed population reaches a lower, but still high, F1-score of $0.81$. This difference is expected because the perturbed class combines galaxies affected by physically distinct mechanisms and therefore covers a wider range of structures than either regular class. Nevertheless, the substantial improvement relative to the two-dimensional diagrams shows that the indices contain complementary information that becomes more effective when considered jointly. The elliptical recovery is also higher than in \citetalias{2026A&A...709A..21S}, possibly because the present reference sample excludes galaxies with stronger visual evidence of disturbance.

\begin{table}
    \centering
    \caption{Classification performance of the calibrated three-class LightGBM model on the independent test sample. The support gives the number of galaxies belonging to each true morphological class.}
    \label{tab:lightgbm_3class_performance}
    \begin{tabular}{lccc}
        \hline
        Class & Precision & Recall & F1-score \\
        \hline
        Elliptical & 0.90 & 0.92 & 0.91   \\
        Spiral     & 0.95 & 0.95 & 0.95  \\
        Perturbed  & 0.82 & 0.81 & 0.81   \\
        \hline
    \end{tabular}
\end{table}

The confusion matrix shows that the main ambiguity is between perturbed galaxies and regular spirals. While $92.5\%$ of ellipticals and $94.9\%$ of spirals are correctly classified, $80.8\%$ of the perturbed systems are recovered. Of the remaining perturbed galaxies, $16.7\%$ are assigned to the spiral class and only $2.6\%$ to the elliptical class. This indicates that many disturbed systems retain a globally disc-like light distribution, even when localised or low-surface-brightness perturbations are present.

\subsection{Mergers \emph{vs.} jellyfish galaxies}
\label{sec:lightgbm_4class}

The three-class experiment tests whether a galaxy is morphologically perturbed, but does not identify the physical mechanism responsible for the disturbance. We now separate the perturbed population into mergers and jellyfish candidates. This four-class experiment provides a multivariate test of the main result from Sect.~4.2: jellyfish candidates form a comparatively coherent morphological population, whereas mergers span a broad and overlapping region of parameter space. The classifier returns four calibrated probabilities, $P_{\rm Ell}+P_{\rm Sp}+P_{\rm Mer}+P_{\rm Jelly}=1$, using the same input indices, training procedure, and independent test sample as the three-class experiment.

The calibrated four-class model reaches an overall accuracy of $0.91$, a balanced accuracy of $0.75$, and a macro-averaged F1-score of $0.76$. The lower balanced accuracy, relative to the overall accuracy, is primarily associated with the merger class. Regular spirals and ellipticals remain well recovered, with F1-scores of $0.95$ and $0.91$, respectively. Jellyfish candidates also reach an F1-score of $0.81$, equal to that of the combined perturbed population in the three-class experiment. Mergers, by contrast, reach a precision of $0.47$, a recall of $0.33$, and an F1-score of $0.39$. The contrast between the jellyfish and merger results confirms that their different degrees of separability are not restricted to the two-dimensional projections. Even when all ten indices are combined, jellyfish candidates retain a comparatively coherent multivariate signature, whereas mergers do not form a compact morphological class. The merger test sample contains only $48$ galaxies, so its class-specific performance estimates have greater statistical uncertainty than those of the other classes. A larger and more representative merger reference sample would improve the precision of these measurements. However, the extensive overlap seen independently in the one-dimensional distributions, two-dimensional diagrams, and full ten-dimensional classifier indicates that sample size alone is unlikely to explain the low merger recovery.

\begin{table}
    \centering
    \caption{
    Classification performance of the calibrated four-class LightGBM
    model on the independent test sample. The support gives the number
    of galaxies belonging to each true morphological class.
    }
    \label{tab:lightgbm_4class_performance}
    \begin{tabular}{lccc}
        \hline
        Class & Precision & Recall & F1-score \\
        \hline
        Elliptical & 0.89 & 0.92 & 0.91 \\
        Spiral     & 0.94 & 0.95 & 0.95 \\
        Merger     & 0.47 & 0.33 & 0.39 \\
        Jellyfish  & 0.81 & 0.81 & 0.81 \\
        \hline
    \end{tabular}
\end{table}

The confusion matrix reinforces this interpretation. The classifier correctly recovers $92.5\%$ of ellipticals, $95.0\%$ of spirals, and $80.6\%$ of jellyfish candidates. The principal ambiguity for jellyfish galaxies is with regular spirals, which account for $13.4\%$ of the true jellyfish sample. This overlap is consistent with the survival of a recognisable stellar disc in many ram-pressure-stripped systems.

Only $33.3\%$ of mergers are assigned to the merger class. Most missed mergers are classified as regular spirals ($45.8\%$), while $14.6\%$ are assigned to the jellyfish class. The dominant confusion with spirals shows that many visually selected mergers or interacting systems retain a disc-like global structure, even when localised tidal features are present.

\begin{figure}
    \centering
    \includegraphics[width=\linewidth]{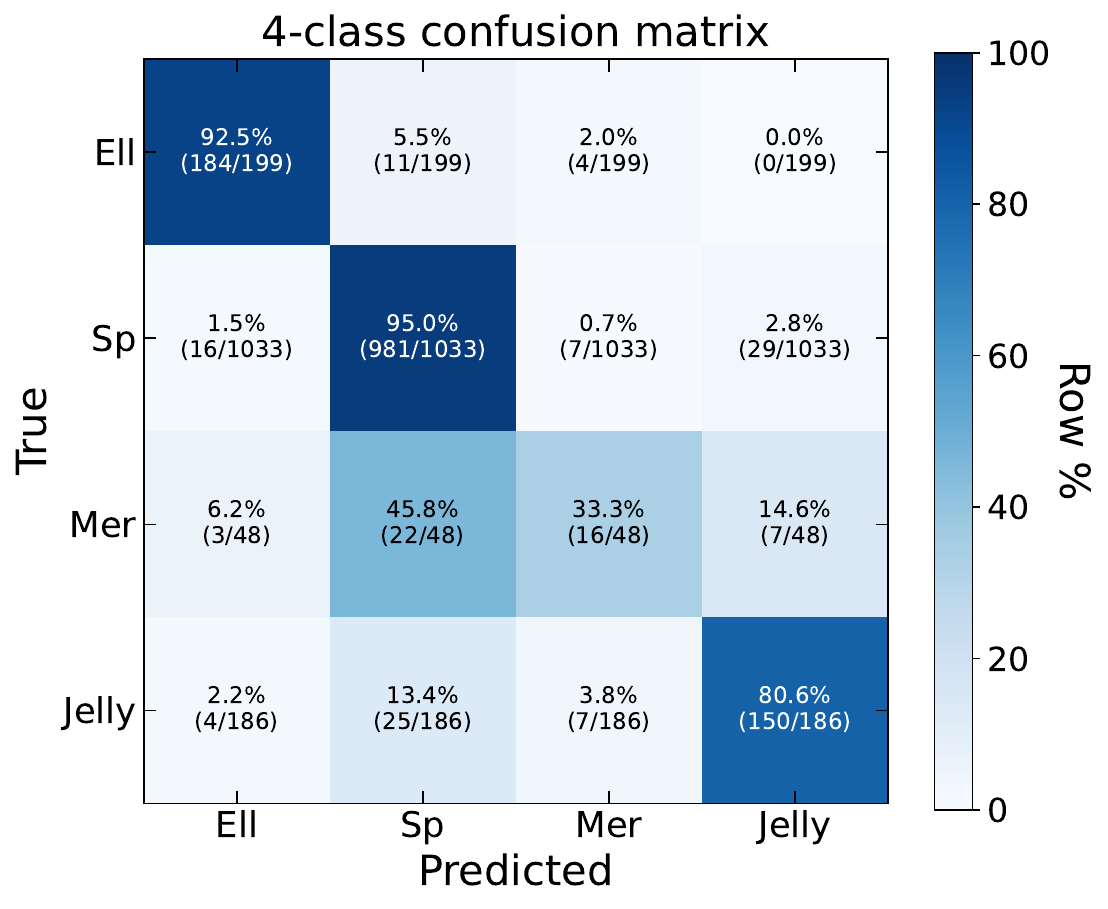}
    \caption{Confusion matrix of the calibrated four-class LightGBM classifier evaluated on the independent test sample. Each row is normalised by the number of galaxies in the corresponding true class. Percentages give the fraction assigned to each predicted class, while the values in parentheses show the corresponding numbers of galaxies.}
    \label{fig:cm_4class}
\end{figure}

Together with the diagnostic diagrams in Sect.~4.2, these results reveal a systematic difference between the two disturbance mechanisms. The optically selected jellyfish candidates share a comparatively repeatable combination of disc structure, asymmetric outer emission, and reduced concentration. Mergers instead represent an evolutionary sequence spanning different interaction stages, mass ratios, gas fractions, orbital configurations, and viewing angles. Their projected light distributions can consequently resemble regular discs, spheroids, jellyfish systems, or highly irregular galaxies. The broad merger distribution is therefore an expected manifestation of the underlying astrophysical diversity of gravitational interactions.

The four-class experiment therefore supports a physical interpretation of the parameter-space structure rather than only a ranking of classification performance. The comparatively high jellyfish recovery reflects the morphological coherence of the optically selected stripping candidates, while the low merger recovery reflects the diversity of gravitational interactions. Projected broad-band morphology can consequently identify many jellyfish candidates, but no single compact morphological signature describes the full merger population.

We apply both calibrated models to the $1,744,454$ galaxies in the parent sample described in Sect.~\ref{sec:parent_sample}. The resulting probabilities provide quantitative measures of morphological similarity to the labelled reference populations. In particular, $P_{\rm Mer}$ should be interpreted with caution: because of the broad merger distribution, small reference sample, and low merger recall, it measures similarity to the visually selected merger class rather than providing a secure physical identification of an ongoing merger.

\section{Conclusions}

We used ten non-parametric morphological indices measured with \textsc{galmex} to study regular spirals, regular ellipticals, mergers, and jellyfish candidates in DESI Legacy Survey imaging. We ranked all two-dimensional combinations of indices using a cross-validated KDE-based classifier and combined the full set of measurements using calibrated LightGBM models. Our main results are as follows.

\begin{enumerate}

    \item The most informative two-dimensional diagrams for separating regular spirals, regular ellipticals, and perturbed galaxies combine Shannon entropy with $M{20}$ or asymmetry-related indices. Entropy measures the complexity of the pixel-intensity distribution, while $M{20}$ and the asymmetry indices add information about the spatial organization of the light. The $E$--$M{20}$ and $E$--$A3$ planes reach balanced accuracies of approximately $0.76$--$0.77$. We use these planes to define probabilistic diagnostic regions, including an uncertain region where the morphological populations overlap.

    \item Jellyfish candidates and mergers show fundamentally different distributions in non-parametric parameter space. Jellyfish candidates occupy a comparatively coherent region characterised, on average, by lower concentration and extended, asymmetric, and off-centre emission. Concentration consequently appears in all of the highest-ranked merger--jellyfish combinations, whose balanced accuracies reach approximately $0.73$. Mergers instead span a much broader structural range, reflecting the diversity of interaction stages, mass ratios, gas fractions, orbital configurations, and viewing angles. The absence of a compact merger locus is therefore an astrophysical result rather than simply a limitation of the classification method.

    \item The full multivariate analysis confirms the structure identified in the two-dimensional diagrams. The calibrated three-class LightGBM model reaches a balanced accuracy and macro-averaged F1-score of $0.89$, showing that broad morphological disturbance can be identified reliably when all ten indices are combined. In the four-class model, jellyfish candidates retain an F1-score of $0.81$, whereas mergers reach a recall of $0.33$ and an F1-score of $0.39$. Most missed mergers are classified as regular spirals, indicating that many interacting systems retain a globally disc-like light distribution despite possessing localised disturbances.

\end{enumerate}

These results show that non-parametric morphology contains information not only about whether a galaxy is disturbed, but also, to a more limited extent, about the physical process responsible for that disturbance. In particular, the comparatively high recovery of jellyfish candidates reflects the repeatable optical signature produced during the visually prominent ram-pressure-stripping phase, whereas the lower merger recovery reflects the intrinsic diversity of gravitational interactions.

The numerical values of the indices and the corresponding diagnostic boundaries depend on image depth, resolution, PSF, segmentation, and preprocessing. The regions and calibrated probabilities presented here are therefore specific to the Legacy/DECam imaging regime and should be recalibrated before being applied to substantially different datasets. We apply the models to the full parent sample of $1,744,454$ galaxies, providing Legacy-specific morphological probabilities for future studies combining wide-field imaging with environmental, photometric, kinematic, or spectroscopic information. Merger probabilities should be interpreted as similarity to the visually selected merger reference sample rather than as secure physical confirmation of an ongoing merger.

\begin{acknowledgements}
The acknowledgements are available in Appendix A.
\end{acknowledgements}

\bibliographystyle{aa} 
\bibliography{bib} 
\begin{appendix}

\section{Acknowledgements}
This research made use of the Python programming language \citep{VanRossum2009} and the packages NumPy \citep{Harris2020},
SciPy \citep{Virtanen2020}, Astropy \citep{Astropy2013,Astropy2018,Astropy2022}, pandas \citep{McKinney2010,Reback2022}, and Matplotlib \citep{Hunter2007}. We gratefully acknowledge the contributions of the volunteers who participated in the Galaxy Zoo and \textit{Fishing for Jellyfish Galaxies} projects through the Zooniverse platform. Their classifications made the construction of the reference samples used in this work possible.

This work has been supported by the Agencia Nacional de Investigación y Desarrollo (ANID) through the Millennium Science Initiative Program NCN2024\_112 (VMS, YLJ, HME); the BASAL project FB210003 (YLJ, HME, AM); the FONDECYT Regular projects 1241426 and 1230441 (YLJ) and 1251882 (AM); and the FONDECYT project 3250511 (CLD). VMS acknowledges additional support from ESO through grant ORP026/2021, and CLD acknowledges support from the ESO Comité Mixto through grant ORP037/2022. AM further acknowledges funding from the HORIZON-MSCA-2021-SE-01 Research and Innovation Programme under the Marie Skłodowska-Curie grant agreement No.~101086388.

\section{GZDECaLS Debiased vs Original Comparison}
\label{app:gzdecals}

\begin{figure}
    \centering
    \includegraphics[width=\linewidth]{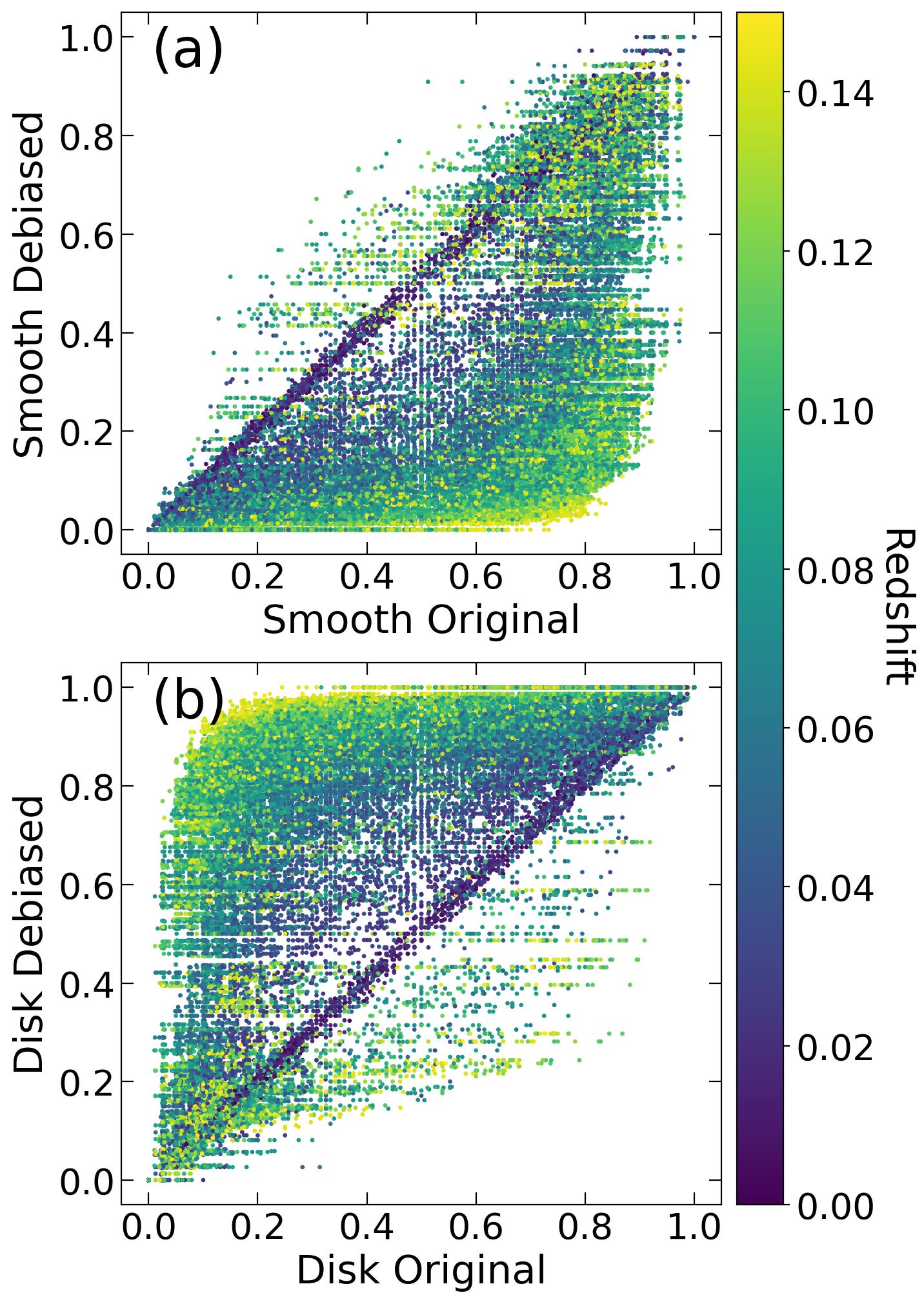}
    \caption{Original versus debiased Galaxy Zoo DECaLS vote fractions for the ``smooth'' [panel (a)] and ``disk/features'' [panel (b)] classifications, with points coloured by redshift. The comparison shows that the debiased vote fractions can differ substantially from the original ones, and that these differences are redshift dependent. This motivates our choice not to use the debiased Galaxy Zoo DECaLS values to define the regular elliptical and spiral reference samples.}
    \label{fig:gz_comparison1}
\end{figure}

\begin{figure*}
    \centering
    \includegraphics[width=\textwidth]{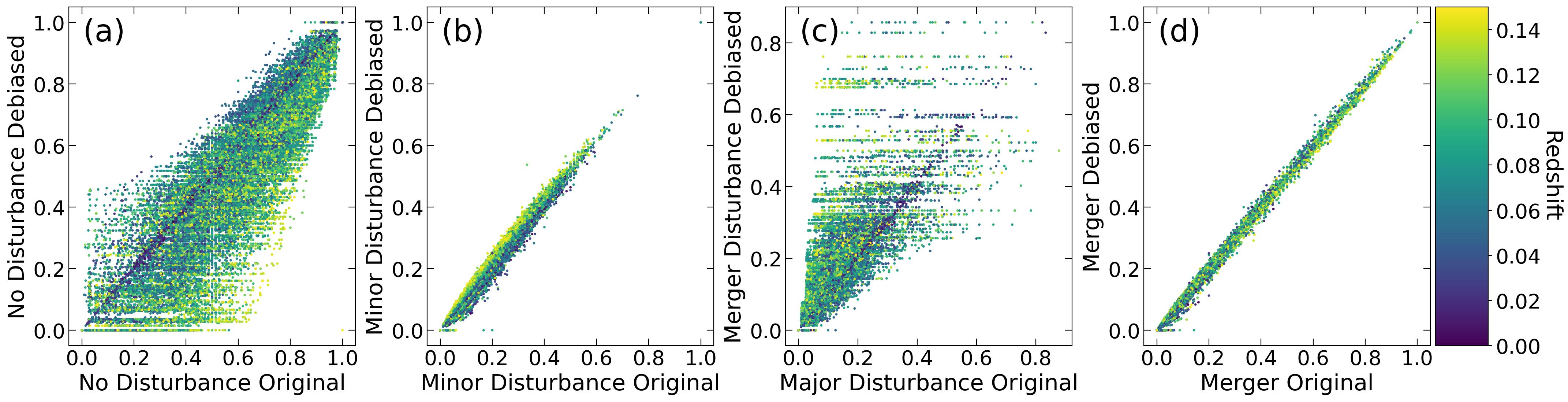}
    \caption{Comparison between the original and debiased Galaxy Zoo DECaLS vote fractions for the disturbance-related classifications: ``No disturbance'' [panel (a)], ``Minor disturbance'' [panel (b)], ``Major disturbance'' [panel (c)], and ``Merger'' [panel (d)]. Points are coloured by redshift. The agreement is better than for the ``smooth'' and ``disk/features'' answers, but redshift-dependent differences are still present, especially for the ``No disturbance'' vote fraction.}
    \label{fig:gz_comparison2}
\end{figure*}

In this appendix we compare the original and debiased Galaxy Zoo DECaLS vote fractions for the classifications that are most relevant to our sample definition. The debiased vote fractions are designed to correct statistical redshift-dependent effects in the visibility of morphological features \citep{2022MNRAS.509.3966W}. However, in this work the labels are used to define high-purity object-by-object reference samples for the calibration of non-parametric morphology. For this purpose, large changes between original and debiased vote fractions can directly affect which galaxies enter the training and validation samples. Figure~\ref{fig:gz_comparison1} shows the comparison for the ``smooth'' and ``disk/features'' classifications, while  Figure~\ref{fig:gz_comparison2} shows the same comparison for the disturbance-related classifications. 

\section{Examples of \texttt{galmex} pre-processing}
\label{app:preproc_examples}
\begin{figure*}
    \centering
    \includegraphics[width=\textwidth]{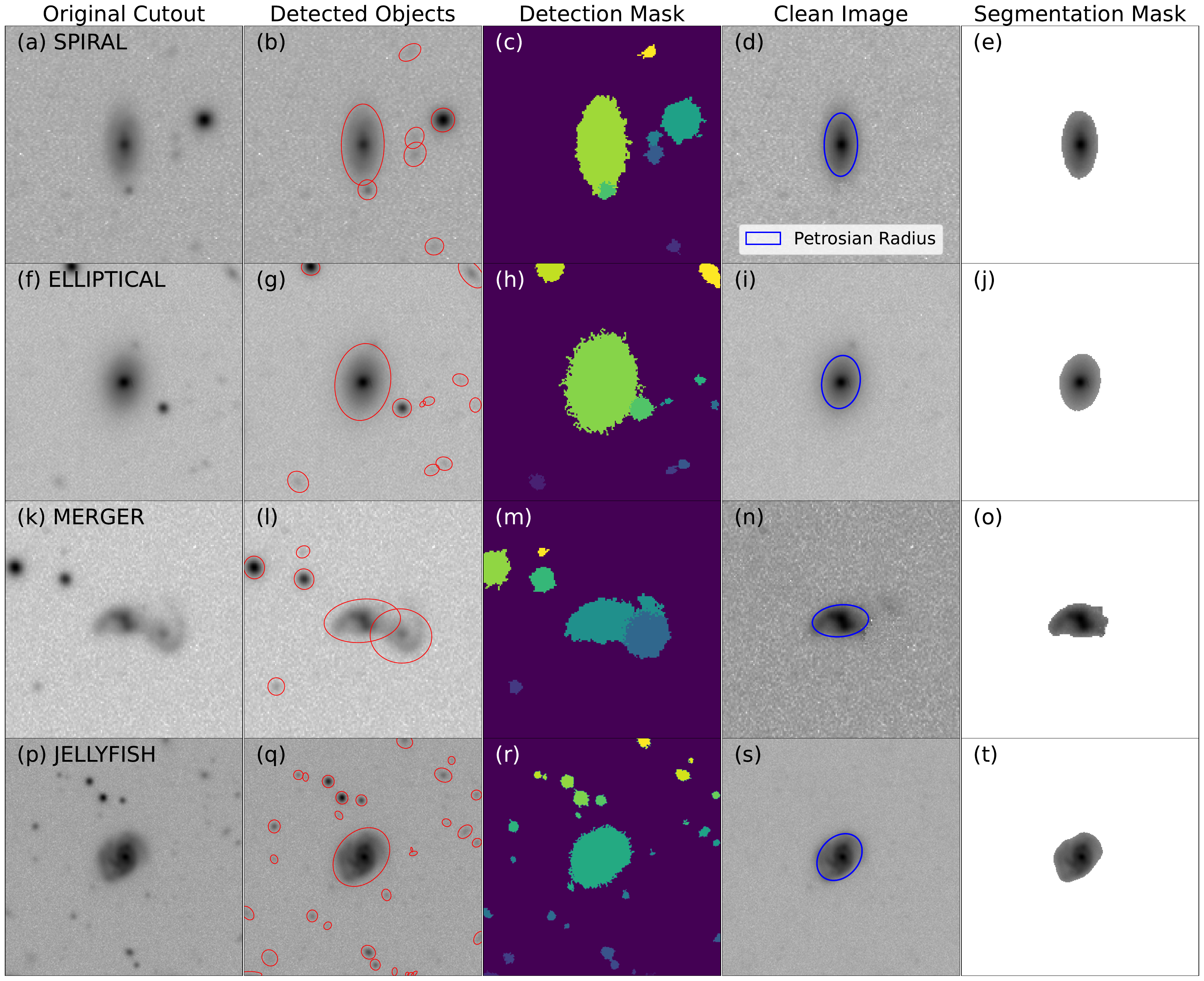}
    \caption{Example of pre-processing for galaxies in the different subsamples. The columns show, respectively, the original cutout, the detected objects, the segmentation mask returned after detection, the clean image, and the segmentation mask used for metric calculation. Each row represent one example for the different morphological classes: spirals (top row), elliptical (second row), merger (third row), and jellyfish galaxy (bottom row). The selected examples consist of panels (b), (g), (n), and (r) from Fig.~\ref{fig:mosaic}.}
    \label{fig:pre_mosaic}
\end{figure*}

In this appendix, we show an example of the \texttt{galmex} pre-processing steps for systems belonging to the different subsamples (regular spirals, regular ellipticals, mergers, and jellyfish galaxies). Fig.~\ref{fig:pre_mosaic} shows the different pre-processing steps, in an analogous way to Fig. F1 in \citetalias{2026A&A...709A..21S}.

\section{Metrics variation with segmentation mask}
\label{app:metrics_variation}

The measurement of non-parametric morphological indicators can depend on the extent of the segmentation mask. To quantify this dependence, we recomputed the segmentation masks and the corresponding morphological metrics while varying the parameter $k$ between 0.5 and 2.5. For each value of $k$, the solid curves show the median value for each morphological class, while the shaded regions indicate the corresponding interquartile range. The vertical black line marks the fiducial value $k=1$ adopted throughout this work.

\begin{figure}
    \centering
    \includegraphics[width=0.8\linewidth]{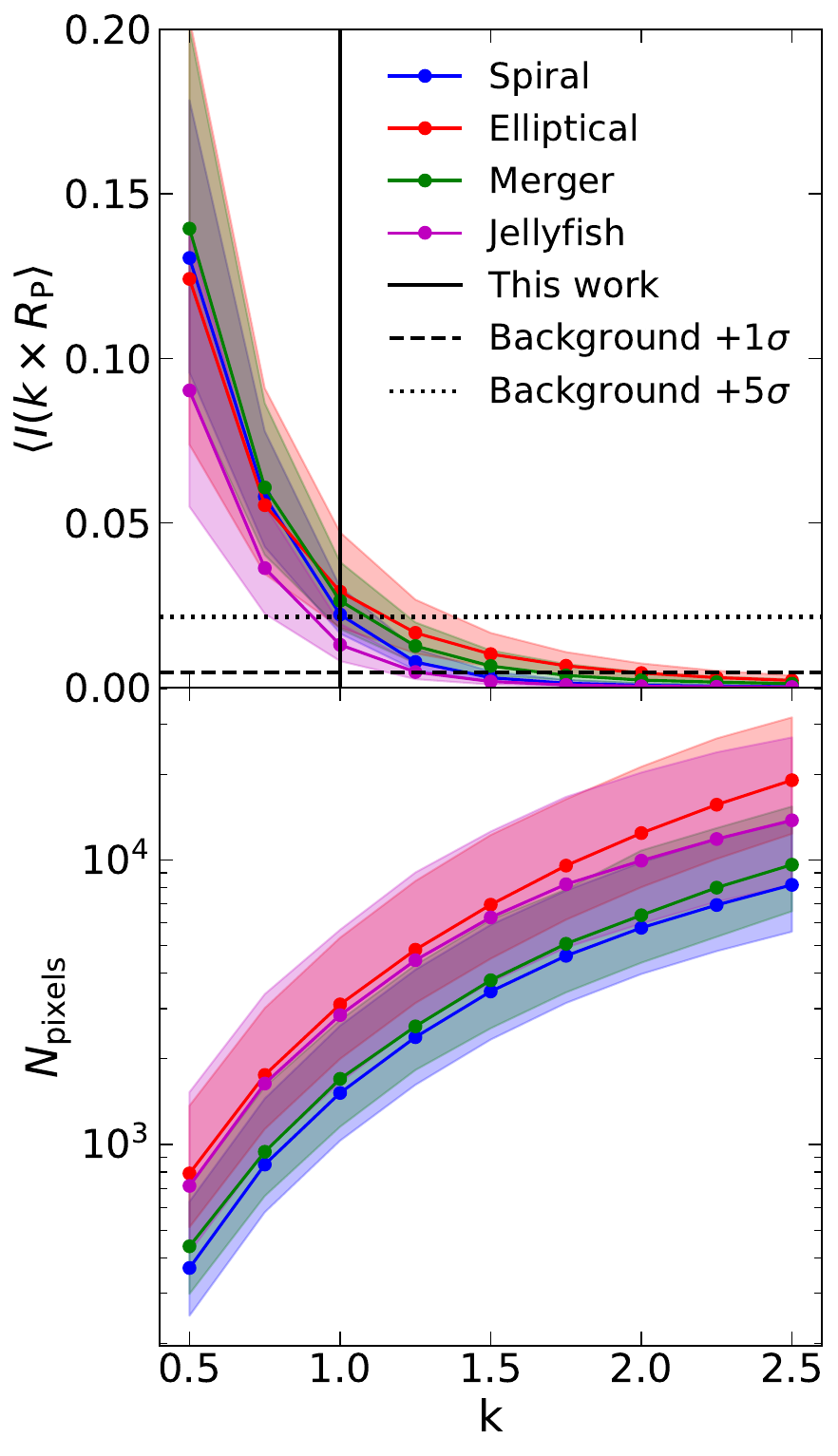}
    \caption{Dependence of the segmentation threshold and mask size on the adopted     value of $k$. The upper panel shows the median intensity threshold evaluated at $kR_{\mathrm{P}}$, while the lower panel shows the corresponding number of pixels included in the segmentation mask. Solid colored curves represent the medians for spiral, elliptical, merger, and jellyfish galaxies, and the shaded regions show their interquartile ranges. The dashed and dotted horizontal lines denote the sample-averaged background-plus-$1\sigma$ and background-plus-$5\sigma$ intensity levels, respectively. The vertical black line marks the fiducial value $k=1$ adopted in this work.}
    \label{fig:kvariation_thresh}
\end{figure}

Figure~\ref{fig:kvariation_thresh} shows how the intensity threshold used to construct the segmentation mask and the resulting number of selected pixels vary with $k$. Increasing $k$ moves the mask boundary toward progressively lower-intensity regions. Consequently, the threshold intensity decreases rapidly, while the number of pixels included in the mask increases monotonically. At small $k$, the masks are restricted to the bright central regions of the galaxies. Conversely, at large $k$, the threshold approaches the background level and an increasing fraction of low-surface-brightness pixels is included.

The horizontal lines in the upper panel indicate the sample-averaged background thresholds $\mathrm{bkg}_{\mathrm{median}}+\sigma_{\mathrm{bkg}}$ and $\mathrm{bkg}_{\mathrm{median}}+5\sigma_{\mathrm{bkg}}$. The fiducial choice $k=1$ corresponds to a threshold comparable to the average $5\sigma_{\mathrm{bkg}}$ level and produces masks containing of order $10^{3}$ pixels. It therefore represents a compromise between retaining the extended galaxy emission and limiting the contribution from background-dominated pixels.

\begin{figure}
    \centering
    \includegraphics[height=0.97\textheight]{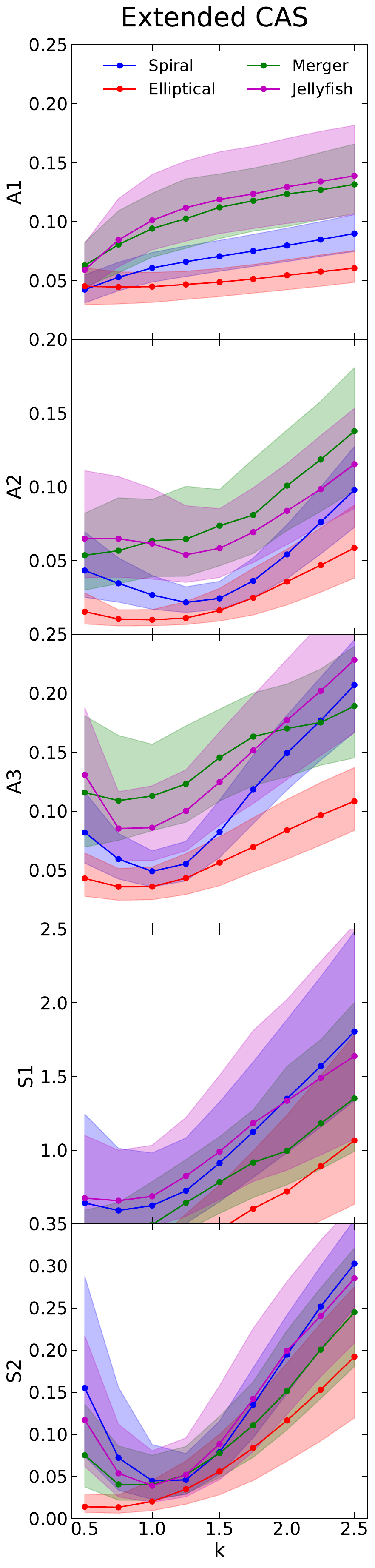}
    \caption{Variation of the Extended CAS metrics with the segmentation-mask parameter $k$.}
    \label{fig:kvariation_cas}
\end{figure}

The Extended CAS indicators exhibit different levels of sensitivity to the segmentation-mask extent, as shown in Figure~\ref{fig:kvariation_cas}. The $A_1$ parameter generally increases with $k$, with the clearest increase occurring for merger and jellyfish galaxies. The higher-order asymmetries, $A_2$ and $A_3$, display a mildly non-monotonic behavior: their median values are approximately constant or decrease slightly around $k\simeq1$, before increasing again.

The smoothness indicators are more sensitive to the outer parts of the segmentation mask. Both $S_1$ and $S_2$ increase toward large $k$, particularly beyond $k\simeq1.25$--1.5. This behavior indicates that low-surface-brightness structures and background fluctuations can make a substantial contribution to the measured small-scale structure when the mask is extended too far. Elliptical galaxies generally retain lower asymmetry and smoothness values than the other populations, although the interquartile ranges show substantial overlap between the different
classes.

\begin{figure}
    \centering
    \includegraphics[height=0.9\textheight]{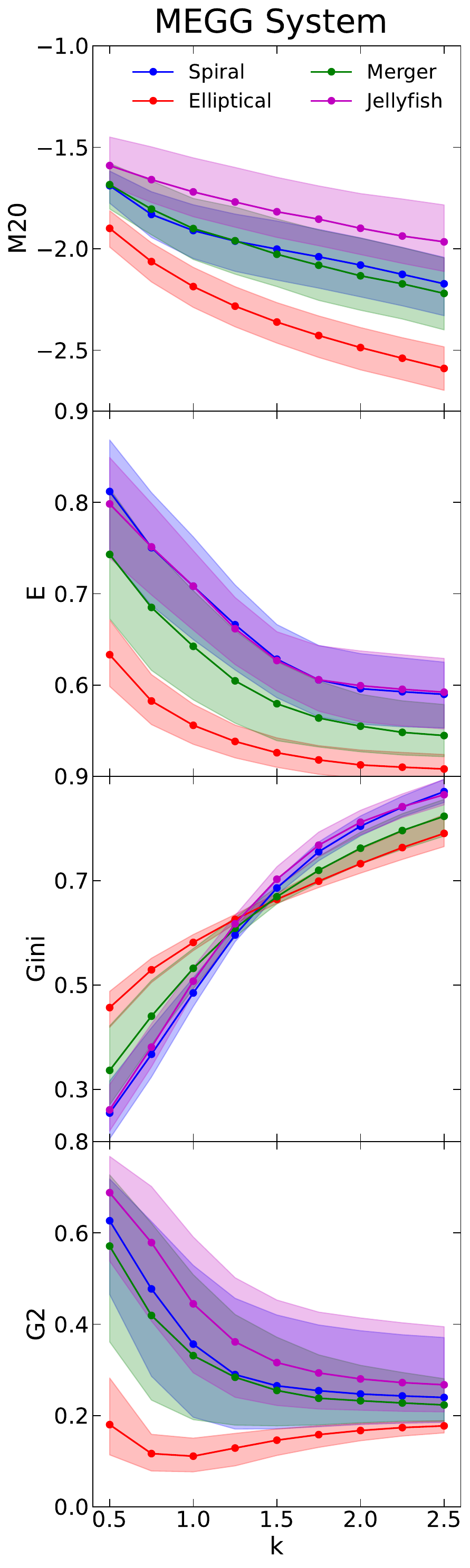}
    \caption{Variation of the MEGG metrics with the segmentation-mask parameter $k$.}
    \label{fig:kvariation_megg}
\end{figure}

Figure~\ref{fig:kvariation_megg} presents the corresponding variation of the MEGG indicators. The $M_{20}$ statistic becomes progressively more negative as $k$ increases, with the strongest variation observed for elliptical galaxies. The entropy $E$ also decreases with increasing mask size and tends to flatten at large $k$. In contrast, the Gini coefficient increases with $k$, reflecting the changing distribution of the total flux as increasingly faint pixels are added to the mask.

For spiral, merger, and jellyfish galaxies, $G_2$ decreases rapidly between $k=0.5$ and $k\simeq1.25$ and subsequently changes more slowly. Elliptical galaxies display a weaker dependence of $G_2$ on $k$, with a small increase toward larger masks. Although the absolute metric values change with the mask definition, the populations continue to exhibit distinct median trends. Nevertheless, the degree of separation and the overlap between their distributions depend on the adopted value of $k$.

Overall, these tests demonstrate that the absolute values of the morphological indicators depend on the definition of the segmentation mask. This dependence is particularly important when low-intensity pixels are included at large $k$. The fiducial choice $k=1$ avoids the most background-dominated regime while retaining emission beyond the bright central regions. Most importantly, the same mask definition is applied consistently to all galaxies, ensuring that the comparison between morphological populations is not driven by differences in the segmentation procedure.

\section{Dependence of results with smoothing bandwidth}
\label{app:smoothing}
\begin{figure}
    \centering
    \includegraphics[width=0.7\linewidth]{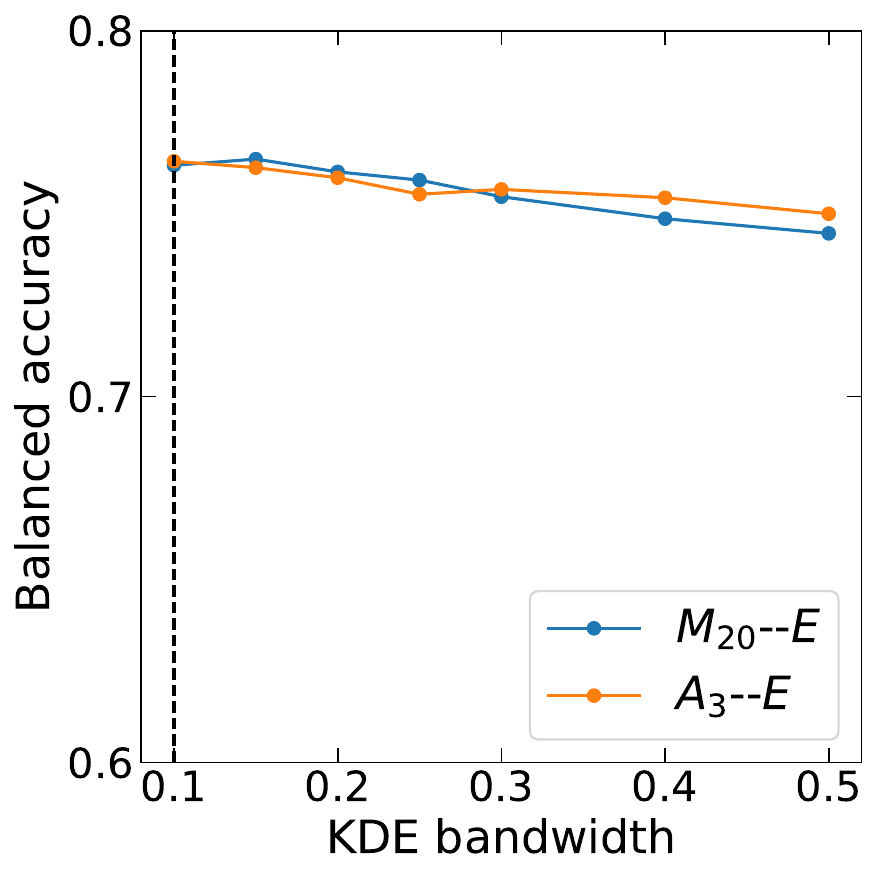}
    \caption{Bandwidth-sensitivity test for the KDE-based classification. The balanced accuracy of the adopted \(E\)--\(M{20}\) and \(E\)--\(A3\) diagnostic planes is shown as a function of the KDE bandwidth \(h\) for the three-class spiral--elliptical--perturbed classification. The small variation across the tested bandwidth range shows that the selected diagrams are not driven by the exact value adopted for \(h\).}
    \label{fig:bandwidth_variation}\label{fig:bandwidth}
\end{figure}

Since galaxies appear as discrete points in the 2-D diagrams combining different non-parametric indexes, we used a Gaussian kernel to smooth the observed distribution and compute the related probabilities. In this appendix we show the stability of the balanced accuracy for the shown combinations ($M20$ vs $E$ and $A3$ vs $E$). Fig.~\ref{fig:bandwidth} shows the variation of the Balanced Accuracy with the selected bandwidth.

\section{Probabilistic combination of the non-parametric indices}
\label{app:lightgbm}

We combine the ten non-parametric morphological indices into calibrated class probabilities using LightGBM classifiers. We consider a three-class experiment separating regular ellipticals, regular spirals, and a combined perturbed population, and a four-class experiment in which mergers and jellyfish galaxies are treated separately. These models complement, rather than replace, the individual morphological diagnostics presented in the main text.

\subsection{Samples, inputs, and data split}
\label{app:lightgbm_samples}
\label{app:lightgbm_features}
\label{app:lightgbm_split}

The classifiers were trained using the visually labelled reference samples described in Sect.~\ref{sec:labeled_sample}. In the three-class experiment, the labels are
\begin{equation}
    y =
    \begin{cases}
        0, & \text{regular elliptical},\\
        1, & \text{regular spiral},\\
        2, & \text{perturbed},
    \end{cases}
\end{equation}
where the perturbed class combines mergers and jellyfish galaxies. The model returns
$(P_{\rm Ell},P_{\rm Sp},P_{\rm Pert})$, with
$P_{\rm Ell}+P_{\rm Sp}+P_{\rm Pert}=1$.

In the four-class experiment,
\begin{equation}
    y =
    \begin{cases}
        0, & \text{regular elliptical},\\
        1, & \text{regular spiral},\\
        2, & \text{merger},\\
        3, & \text{jellyfish},
    \end{cases}
\end{equation}
and
$P_{\rm Ell}+P_{\rm Sp}+P_{\rm Mer}+P_{\rm Jelly}=1$.
Objects with missing or non-finite values in any input index were excluded, removing only three jellyfish galaxies. The resulting sample sizes are listed in Table~\ref{tab:lightgbm_samples}.

\begin{table}
    \centering
    \caption{Composition of the development and independent test samples. The perturbed class in the three-class experiment combines mergers and jellyfish galaxies.}
    \label{tab:lightgbm_samples}
    \begin{tabular}{ccccc}
        \hline
        Experiment & Class & Total & Development & Test \\
        \hline
        3-class
        & Elliptical & 996  & 797  & 199  \\
        & Spiral     & 5164 & 4131 & 1033 \\
        & Perturbed  & 1169 & 935  & 234  \\
        \hline
        4-class
        & Elliptical & 996  & 797  & 199  \\
        & Spiral     & 5164 & 4131 & 1033 \\
        & Merger     & 240  & 192  & 48   \\
        & Jellyfish  & 929  & 743  & 186  \\
        \hline
    \end{tabular}
\end{table}

For each galaxy, the input vector is
\begin{equation}
    \boldsymbol{x}
    =
    \left(
    C,\,
    A_1,\,
    A_2,\,
    A_3,\,
    S_1,\,
    S_2,\,
    M_{20},\,
    E,\,
    G,\,
    G_2
    \right).
\end{equation}
No photometric, spectroscopic, colour, mass, redshift, or environmental quantities were supplied to the classifiers.

Each reference sample was divided using a stratified split into a development sample containing 80\% of the galaxies and an independent test sample containing the remaining 20\%. The development sample was used for training and probability calibration, whereas the test sample was reserved exclusively for the final evaluation. This separation is particularly important because regular spirals constitute approximately 70\% of the sample, while mergers represent only approximately 3\% of the four-class sample.

\subsection{Training and probability calibration}
\label{app:lightgbm_smote}
\label{app:lightgbm_classifier}
\label{app:lightgbm_calibration}

Class imbalance within the development sample was treated using the Synthetic Minority Over-sampling Technique (SMOTE), with five neighbours. SMOTE generates intermediate points between neighbouring objects of the same minority class in the ten-dimensional morphological parameter space. It was applied only to the training portion of each cross-validation fold; neither the corresponding validation fold nor the independent test sample was used to generate synthetic objects.

The classes were modelled using a LightGBM gradient-boosted decision-tree classifier. The same configuration was adopted for the three- and four-class experiments, except for the number of output classes, as summarised in Table~\ref{tab:lightgbm_configuration}.

\begin{table}
    \centering
    \caption{Configuration of the LightGBM classifiers and probability calibration.}
    \label{tab:lightgbm_configuration}
    \begin{tabular}{lc}
        \hline
        Parameter & Adopted value \\
        \hline
        Number of input indices & 10 \\
        Number of classes & 3 or 4 \\
        Development fraction & 0.80 \\
        Independent test fraction & 0.20 \\
        Number of boosting trees & 1000 \\
        Learning rate & 0.03 \\
        Maximum number of leaves & 31 \\
        SMOTE neighbours & 5 \\
        Calibration folds & 5 \\
        Calibration transformation & Sigmoid \\
        Random seed & 42 \\
        \hline
    \end{tabular}
\end{table}

For each galaxy, the preferred class is the one with the largest probability,
\begin{equation}
    \widehat{y}_i
    =
    \underset{c}{\operatorname{arg\,max}}\,
    P_{i,c},
\end{equation}
although the complete probability vector is retained to quantify classification confidence.

Because probabilities returned directly by boosted decision trees are not necessarily calibrated, we apply a sigmoid transformation using five-fold stratified cross-validation within the development sample. In each fold, the complete SMOTE--LightGBM pipeline is fitted to four-fifths of the development sample and used to predict the remaining fifth, producing out-of-fold probabilities for every development galaxy. The sigmoid calibration is fitted to these predictions, after which the base classifier is trained on the complete development sample. In the multiclass case, calibration is performed independently for each class in a one-versus-rest scheme and the probabilities are renormalised to sum to unity.

The calibrated probabilities have a frequency-based interpretation for samples comparable to the reference data. For example, among galaxies assigned $P_{\rm Jelly}\simeq0.8$, approximately 80\% are expected to belong to the jellyfish reference class. Both the raw and calibrated models are evaluated on the same independent test sample.

\subsection{Performance and probability diagnostics}
\label{app:lightgbm_metrics}
\label{app:lightgbm_probability_metrics}

We report the precision, recall, F1-score, and support of each class, together with the overall accuracy. Because of the unequal class sizes, we additionally use the balanced accuracy, defined as the mean class recall, and the macro-averaged F1-score, defined as the mean class F1-score. These quantities give equal weight to all morphological classes and are therefore more informative than the overall accuracy alone.

The numerical probabilities are assessed using the multiclass logarithmic loss,
\begin{equation}
    \mathcal{L}_{\rm log}
    =
    -\frac{1}{N}
    \sum_{i=1}^{N}
    \ln P_{i,y_i},
\end{equation}
where $P_{i,y_i}$ is the probability assigned to the true class of galaxy $i$, and the multiclass Brier score,
\begin{equation}
    {\rm BS}
    =
    \frac{1}{N}
    \sum_{i=1}^{N}
    \sum_c
    \left(
        P_{i,c}-Y_{i,c}
    \right)^2,
\end{equation}
where $Y_{i,c}=1$ when galaxy $i$ belongs to class $c$ and is zero otherwise. Lower values indicate more accurate probability estimates. Both statistics are also calculated for a class-prior baseline that assigns all galaxies the class fractions measured in the development sample.

Calibration is inspected by grouping galaxies into fixed probability intervals and comparing the mean predicted probability with the observed class fraction. We also examine the maximum probability,
$P_{{\rm max},i}=\max_c P_{i,c}$, by comparing its mean value with the observed accuracy of the preferred classification. A binned expected calibration error is used as a supplementary diagnostic, together with the logarithmic loss, Brier score, and class-specific statistics. Finally, thresholds in $P_{\rm max}$ are used to quantify the trade-off between sample completeness and classification reliability.

\subsection{Interpretation and limitations}
\label{app:lightgbm_limitations}

The inferred probabilities are conditional on the labelled reference samples and on the observational properties of the images used to measure the indices. They should not be interpreted as universal probabilities independent of survey depth, angular resolution, signal-to-noise ratio, rest-frame wavelength, redshift, or segmentation procedure. Probability calibration may also require reassessment when the classifier is applied to a population with substantially different intrinsic class fractions.

In the three-class model, $P_{\rm Pert}$ measures membership in the combined merger--jellyfish population and does not identify the physical origin of the disturbance. The four-class model attempts this separation, but its merger probabilities are limited by the diversity of merger morphologies and the comparatively small merger reference sample. Within these limitations, the calibrated probabilities provide a compact representation of the multidimensional morphology and a quantitative measure of confidence in the preferred classification.

\end{appendix}

\end{document}